\documentclass[a4paper,11pt]{article}
\pdfoutput=1
\usepackage{cite}
\usepackage{jheppub}
\usepackage{hyperref}
\usepackage{color,amsmath}
\usepackage{comment}
 
\hypersetup{colorlinks=true}
\usepackage[T1]{fontenc} 
\usepackage[normalem]{ulem}
\title{Non-relativistic Strings: Classical solutions and exactly solvable models  }
\author{Aritra Banerjee$^1$, Adrita Chakraborty$^2$, Nibedita Padhi$^3$ }
\affiliation{$^1$Birla Institute of Technology and Science, Pilani Campus, Pilani, Rajasthan 333031, India.}
\affiliation{$^2$Faculty of Physics and Applied Computer Science, AGH University of Krakow, al. A. Mickiewicza 30, 30-059 Krakow, Poland.}
	 \affiliation{$^3$Department of Physics, Indian Institute of Technology Kharagpur-721302, India.}
     \emailAdd{aritra.banerjee@pilani.bits-pilani.ac.in}
	\emailAdd{achakraborty@agh.edu.pl}
	 \emailAdd{nibedita.phy@iitkgp.ac.in}
\abstract{We discuss classical closed string solutions in non-relativistic two-sphere target spaces. These classes of solutions closely relate to the GKP-type, spinning and pulsating strings for the relativistic case. We derive the string dynamics in each case and construct relevant dispersion relations, both from the string Newton-Cartan intrinsic sigma model and using a large speed of light expansion of relativistic Polyakov action. We further discuss construction and characteristics of exactly solvable Neumann-Rosochatius-like dynamical systems corresponding to strings in leading and subleading orders of the expanded Polyakov theory.

}


\begin{document}
\maketitle
	\flushbottom
	\section{Introduction} 
    \label{introduction}

The paradigm of high-energy physics is inherently relativistic. From particles to strings, theories are designed in a manifestly Lorentz covariant way since that seems to be the proclaimed symmetry of physics at high energies. This is counterintuitive to our mundane experiences, where we live in a non-Lorentzian, more specifically, a Galilean world. Hence a non-relativistic approximation to relativistic physics is often the need of the hour, for a smooth interpolation between these two experiences. Most recently, there has been a resurgence in studying non-relativistic theories, fuelled by non-Lorentzian versions of holographic dualities \footnote{See \cite{Bergshoeff:2022eog} for a recent review of progress along this direction.}. This has in turn started the need to better understand similar limits of relativistic string theories, and as the final frontier, to study non-Lorentzian theories of quantum gravity. 
\medskip

In fact, the non-relativistic version of general relativity has been understood very well for some time now, where one writes the theory in a covariant formalism on a \textit{Newton-Cartan manifold} \cite{cartan1986manifolds, HAVAS:1964vek} instead of a Lorentzian one. The physics in Newton-Cartan setting has been revived in connection with Non-Lorentzian systems recently \cite{Hartong:2022lsy} and it gives an elegant geometric understanding of a world with an infinite speed of light.  The covariant formulation has numerous applications, including in constructing field theories with Galilean symmetries or the conformal cousin thereof\footnote{See for example \cite{Son:2013rqa, Hartong:2015wxa, VandenBleeken:2015rzu, Festuccia:2016caf,Bagchi:2022twx,Lambert:2024yjk} and references therein. The list is of course far from complete.}, and even in the case of Post-Newtonian (PN) approximations to General Relativity \cite{Hartong:2023ckn}.
\medskip

 In another related avenue of progress, non-relativistic string theory has become a promising candidate to explore quantum gravity in non-Lorentzian target spacetimes\footnote{A nice introduction can be found in \cite{Oling:2022fft}.}. Such theories were originally studied by Gomis and Ooguri  \cite{Gomis:2000bd} (See also \cite{Danielsson:2000gi, Klebanov:2000pp, Danielsson:2000mu}) as a consistent, unitary, UV finite theory on flat spacetime (having a compact direction) with a near-critical Kalb-Ramond field. The spectra of various string states as well as the spacetime S-matrix of these theories follow a definite non-relativistic symmetry. The non-relativistic string formalism was further extended to general spacetime in \cite{Gomis:2005pg}. Being UV finite, these theories allow suitable quantization of the non-relativistic spacetime geometry. However, 
 unlike relativistic string theories, the nonlinear sigma model for a probe non-relativistic string defines a 2D relativistic worldsheet theory on a non-relativistic target space. These non-relativistic target manifolds are also conveniently equipped with Newton-Cartan(NC) spacetime geometry.
 \medskip
 
 Such Newton-Cartan (NC) geometries are generally formalized in two settings, namely,  torsional Newton-Cartan (tNC) geometry 
   and stringy Newton-Cartan (sNC) geometry. The tNC geometry was first put forth in \cite{Christensen:2013lma, Christensen:2013rfa} and are further studied extensively in \cite{Hartong:2014oma, Festuccia:2016caf, Harmark:2017rpg, Harmark:2018cdl, Gallegos:2019icg,  Oling:2022fft}. In particular, these are utilized to construct a novel covariant string action by performing target space null reduction in the Polyakov action of a fundamental string \cite{ Harmark:2017rpg}, thus delivering a promising platform to study non-relativistic holography. The latter setting was originally developed in \cite{Andringa:2012uz}, where authors introduce a unique geometric constraint on the target space, where time and one of the distinguished space directions are singled out from the rest, which effectively separates the resulting non-Lorentzian target space into longitudinal and transverse directions. This makes it somewhat different from the usual Riemannian geometry. 
\medskip
   
   In \cite{Bergshoeff:2018yvt}, these geometries are efficiently implemented to study non-relativistic string theory on curved backgrounds. Moreover, sNC geometry is found in \cite{Bergshoeff:2019pij} as the limit of the Riemannian geometry of general relativity with fluxless 2-form fields. More interesting approaches based on this framework can be found in \cite{Bergshoeff:2018vfn,  Kluson:2018egd, Bergshoeff:2021bmc, Yan:2021lbe, Bergshoeff:2019pij, Gomis:2019zyu}.  Currently, there exist two equivalent approaches for a potential analysis of non-relativistic strings starting from a relativistic string theory \cite{Harmark:2019upf}. Among these, one undergoes the reduction of a relativistic string along a null isometry in target space whereas the other one involves large $c$ expansion of the target space geometry, $c$ being the speed of light \cite{Hartong:2021ekg, Hartong:2022dsx}. Here one considers $1/c^2$ expansion of all the fields in the Polyakov action. These expansions can further be of two types, the \textit{particle} ones and the \textit{string} ones. In the particle $1/c^2$ expansion, only time direction is singled out, and the metric leads to a tNC or an ordinary NC geometry. In contrast, string $1/c^2$ expansions take care of the transverse-longitudinal split\footnote{Similar string small $c$ expansions can be performed for the opposite limit, for the \textit{Carrollian} target spaces (like near horizon limit of Black Holes), and more details can be found in \cite{Bagchi:2023cfp, Bagchi:2024rje}.}. In fact, the next-to-leading order action in the string expansion happens to be exactly the same as the known action of the intrinsic sNC non-relativistic string theory.
   \medskip

   Worldsheet theory of non-relativistic strings moving in non-Lorentzian geometries has its own significance in diverse aspects of non-AdS holography \cite{Christensen:2013lma, Hartong:2014oma, Fontanella:2024rvn}, since it is based on non-Riemannian geometry while the conformal symmetry on the worldsheet remains invariant. However the exact formulation is far from complete. Very recently, people have also discussed the role of these string theories within a larger web of BPS decoupling limits related by stringy dualities \cite{Blair:2023noj, Gomis:2023eav, Blair:2024aqz}. Connections to double field theory \cite{Park:2020ixf, Blair:2020gng}, supersymmetric generalizations \cite{Gomis:2005pg,Blair:2019qwi,Bergshoeff:2022pzk} and Matrix theories \cite{Danielsson:2000gi, Blair:2024aqz} (see also \cite{Guijosa:2025mwh}) have all been explored, with the list of references being nowhere near exhaustive.
   \medskip
   
   However there is a less explored realm in the land of non-relativistic strings and associated holography, and that stems from the question of integrability\footnote{For related review of integrability in relativistic (AdS) holography case, reader can have a look at \cite{Beisert:2010jr, Zarembo:2017muf}.}. The classical integrability of $AdS$ strings is a powerful motivation to find the fate of the same for non-relativistic sigma models. Some work along the lie algebraic point of view has been done \cite{Kluson:2017ufb, Roychowdhury:2019vzh}, but the dust has not settled on the question yet. A related avenue of enquiry is that of classical string solutions \cite{Plefka:2005bk}, since the power of integrability in relativistic holography makes sure that semiclassical closed string states on $AdS_5\times S^5$, described by solitonic configurations, can be used to compute the dimensions of the corresponding super-Yang-Mills (sYM) gauge-theory operators on the boundary.
   Consequently, series of works have appeared on various string and brane solutions in the non-relativistic framework \cite{Roychowdhury:2019olt, Roychowdhury:2020dke, Roychowdhury:2020yun, Roychowdhury:2021wte, Roychowdhury:2019sfo, Roychowdhury:2020abu, Roychowdhury:2020cnj} where the presence of integrable features in the non-relativistic setup has been explored. Some generic classes of string ansatz, for instance, closed and rigid rotating and pulsating strings are found to apparently reduce the physical system into integrable dynamical models. Such integrable models are closely related to the class of Neumann-Rosochatius ones \cite{Arutyunov:2003uj,Arutyunov:2003za}, where the motion of a harmonic oscillator on a unit sphere/AdS under the influence of an extra inverse-square-type potential is considered to map onto generic rotating and pulsating probe strings in various compact relativistic target spaces \cite{Ahn:2008hj, Hernandez:2014eta, Hernandez:2015nba, Arutyunov:2016ysi, Hernandez:2017raj, Hernandez:2018gcd, Hernandez:2018lvh, Nieto:2018jzi, Chakraborty:2019gmt, Hernandez:2020igz, Chakraborty:2020las, Chakraborty:2022iuk, Chakraborty:2022eeq, Hernandez:2022dmf}.
\medskip 

Motivated by this, in this article, we will investigate the solutions and the solvable features of non-relativistic string sigma model with Lorentzian worldsheet by choosing the target space as non-relativistic cousin of $\mathbb{R}\times S^2$. Firstly, we use the intrinsically non-relativistic string action constructed from a sNC formalism, and study well known analogues of rigid spinning solutions. In the second one, as described before, we take a probe relativistic fundamental string and perform a non-Lorentzian expansion of the target space vielbeins at the limit $c\rightarrow \infty$. For this scenario, we choose to embed, from relativistic intuition, circular spinning and pulsating strings, albeit order by order. At definite orders we reduce the worldsheet Lagrangians into those of solvable models containing harmonic oscillator-type and inverse-square-type potentials akin to the integrable Neumann-Rosochatius model. 
\medskip    

     When we consider the non-Lorentzian scenario for an intrinsically non-relativistic spinning string, the sigma model does not have any standard kinetic term, energy-spin dispersion relations have different structures and the nature of the longitudinal solution is imposed by auxiliary fields. Furthermore, the $\frac{1}{c^2}$-expanded target space, probed by relativistic fundamental spinning and pulsating string, presents exactly solvable dynamics from both the leading order and next-to-leading order Lagrangians and Hamiltonians. In this case, all solvable models acquire potentials similar to the Neumann-Rosochatius (NR) integrable framework, but the usual kinetic terms remain elusive. Nevertheless, the embedding fields in the leading-order and subleading-order are highly constrained among each other, which causes a sufficient reduction in the number of degrees of freedom in the dynamics. We further quantize the total energy of these exactly solvable systems order by order using the Bohr-Sommerfeld quantization principle. For spinning string, we assume an arbitrary quantum number to designate the discrete energy states whereas the energies of different pulsating string states are expressed in terms of the usual adiabatic oscillation numbers in different orders.
     \medskip
     
The rest of the article is organized as follows. In section \eqref{section 2}, we discuss the string worldsheet Lagrangian for the stringy Newton-Cartan sigma model for a GKP-like and a rigid spinning string on the intrinsically non-relativistic target space $\mathbb{R}\times S^2$. The dispersion relation for conserved charges in the GKP case turns out to be comparable to the relativistic version thereof, while the spinning string gives rise to a dispersion relation akin to the small momentum limit of Giant Magnon \cite{Hofman:2006xt} solution.  In section \eqref{section 3}, this is followed by the construction of stringy Newton-Cartan expansion of $\mathbb{R}\times S^2$ target space geometry in orders of $\frac{1}{c^2}$ in the $c\rightarrow \infty$ limit. We then study the dynamics of probe spinning string in this sNC target space in section \eqref{section 4} where we present the reduction of both leading order and subleading order of the non-Lorentzian worldsheet actions as some Neumann-Rosochatius type exactly solvable models with specific constraints. Moreover, we use the exact solutions of these models to quantize the total energy by using Bohr-Sommerfeld quantization rule. Section \eqref{section 5} is devoted to the similar studies for leading and next-to-leading order dynamics of pulsating string embedding and corresponding quantization in terms of adiabatic oscillation number. We conclude our article with some potential future prospects in section \eqref{section 6}. Appendices contain extra details and discussions. 

\section{Worldsheet Lagrangian for non-relativistic string in $\mathbb{R} \times S^2$} \label{section 2}
\subsection{The formalism}
In this section, we will discuss a probe fundamental string in a string Newton-Cartan cousin of $\mathbb{R}\times S^2$ target space. We are interested in carrying out an intrinsic stringy Newton-Cartan formalism as discussed in \cite{Bidussi:2021ujm} for the action of a string with relativistic worldsheet, assumed to be spinning (or pulsating) in the chosen non-relativistic target space. The action for the sigma model of non-relativistic string theory on an arbitrary string Newton-Cartan geometry
\cite{Bergshoeff:2018yvt} is given by,
\begin{align}
\label{masteraction}
   S_{P}^{NR}=-\frac{T}{2}\int d^2\sigma \Bigg[\sqrt{-h}h^{\alpha \beta}\partial_{\alpha}X^{\mu}\partial_{\beta}X^{\nu}H_{\mu\nu}+\epsilon^{\alpha \beta} (\xi e_{\alpha}\tau_{\mu}+\bar{\xi} \bar{e}_{\alpha}\bar{\tau}_{\mu})\partial_{\beta}X^{\mu}
   \Bigg]
\end{align}
This action can be constructed from the relativistic Polyakov one (with a NS-NS flux) by considering the correct contraction procedure for the coordinates \cite{Fontanella:2021hcb}. The pre-contraction vielbein expands in a large $c$ limit where longitudinal and transverse coordinates appear on different footing:
\begin{equation}
e_{\mu}^{a}=c \tau^{a}_{\mu}+\frac{1}{c}m^{b}_{\mu}\epsilon^{a}_{b}+\mathcal{O}(c^{-2}),     ~~~e_{\mu}^{a'}=E_{\mu}^{a'} + \mathcal{O}(c^{-1})
\end{equation}
Here
$a = 0,1 $ denotes the longitudinal directions and $a'= 2...D$ denotes the transverse directions of the manifold's tangent space respectively. Greek indices are usual target space directions. 
Without going into further subtleties, in the action above, we can construct metrics along the two subspaces:
\begin{equation}
H_{\mu\nu}=E_{\mu}^{a'}E_{\nu}^{b'}\delta_{a'b'}+(\tau_{\mu}^a m_{\nu}^b+\tau_{\nu}^a m_{\mu}^b)\eta_{ab}, \quad \tau_{\mu\nu} = \tau_\mu^{~a} \tau_\nu^{~b}\eta_{ab} 
\end{equation}
where  
the gauge fields $\tau_{\mu}^a$, $m_{\mu}^a$ and $E_{\mu}^{a'}$ define the string Newton-Cartan geometry. The flat metrics in these two directions are given by $\delta_{a'b'}= (0,0,1...1)$ and $\eta_{ab} = (-1,1,0,..0)$. Further, $\xi$ and $\bar{\xi}$ represent the worldsheet coordinate dependent Lagrange multipliers, which one has to introduce in the contraction procedure. 
Again, the longitudinal and transverse vielbein fields have to satisfy constraint relations among themselves and these are given by 
\begin{align}
    D_{[\mu}\tau_{\nu]}^{a}=0,\quad E_{\mu}^{a'}E_{b'}^{\mu}=\delta_{b'}^{a'},\quad \tau^{\mu}_{a}\tau^{b}_{\mu}=\delta^{b}_{a},\quad\\
    \tau_{\mu}^{a}\tau^{\nu}_{a}+ E_{\mu}^{a'}E_{a'}^{\nu}=\delta^{\nu}_{\mu},\quad \tau^{\mu}_{a}E_{\mu}^{a'}=0,\quad \tau_{\mu}^{a}E^{\mu}_{a'}=0.
\end{align}
Additionally, the one-forms $\tau_{\mu}$ and $\bar{\tau}_{\mu}$ can be expressed as linear combinations of $\tau_{\mu}^{0}$ and $\tau_{\mu}^{1}$, specifically as $\tau_{\mu}^{0} \pm \tau_{\mu}^{1}$ in the light-cone coordinates, respectively. Similarly, the worldsheet vielbein fields $e_{\alpha}, \bar{e}_\alpha$ are expressed as $e_{\alpha}^{0}\pm e_{\alpha}^{1}$.
\medskip

In this work, we are interested in a $\mathbb{R} \times S^2$ target space, given the Lorentzian geometry with a radius $R$:
\begin{equation}
    ds^2=-R^2dt^2+R^2( d\theta^2+\sin^2{\theta}d\phi^2),
\end{equation}
To take the string Newton-Cartan (sNC from this point onward) limit, we choose the coordinates $t$ and $\theta$ as the longitudinal directions and $\phi$ as the transverse direction. Let us now write various sNC data. Note that the relativistic vielbeins, scaled appropriately by a large $c$ are given by:
\begin{eqnarray}
  &&  e^{\hat{0}}=e^{\hat{0}}_{t}dt=e^{0}_{t}dt=Rdt\nonumber\\&&
     e^{\hat{1}}=e^{\hat{1}}_{\theta}d\theta=e^{1}_{\theta}d\theta=Rd\theta\nonumber\\&&
     e^{\hat{2}}=e^{\hat{2}}_{\phi}d\phi=E^{2}_{\phi}d\phi=R\sin{\theta}d\phi
\end{eqnarray}
Now we introduce the non-relativistic contraction of the coordinates as 
\begin{equation}
    R=cl,~t=\mathfrak{t},~\theta={\vartheta},~\phi=\frac{\varphi}{c}, ~~c\to \infty
\end{equation}
where the length scale connects to the tension $l=\sqrt{\alpha'_{NR}}(\lambda_{NR})^{1/4}$, where $\alpha'_{NR}$ is the string length square and $\lambda_{NR}$ is the associated 't Hooft coupling.  For simplicity, we will set $l=1$ in what follows, keeping only the effective tension to set the string scale. Using the above scaling of coordinates we find the sNC vielbiens
\begin{eqnarray}
   \tau^{0}_{t}=1,~~\tau^{1}_{\theta}=1,~~  E^{2}_{\phi}=\sin{\vartheta}
    \label{2.7}
\end{eqnarray}
Next we choose to work in the conformal gauge, since our worldsheet is still manifestly relativistic, by fixing the worldsheet metric to be the flat one: $h_{\alpha \beta}=\eta_{\alpha \beta}$. This allows us to fix the worldsheet vielbien fields as  $e_{0}^{0}=e_{1}^{1}=1$ in light-cone coordinates, resulting in:
\begin{eqnarray}
    \mathcal{L}_{P}^{NR}=-\frac{T}{2}\left[\sin^2\vartheta(\varphi'^2-\dot{\varphi}^2)+[(\xi+\bar{\xi})({\mathfrak{t}'}-\dot{\vartheta})+(\xi-\bar{\xi})(\vartheta'-\dot{\mathfrak{t}})]\right].
    \label{Lagrangian 1}
\end{eqnarray}
We also scale the string tension as $T\to c^2T$ here. To distinguish the notation, dots and primes now on would correspond to derivatives w.r.t $\tau$ and $\sigma$.
Note the Lagrange multipliers in the action can also be thought of as having a light-cone form, so that $\xi \pm \bar{\xi} = \xi_{0,1}$.
From the above Lagrangian, we have the constraint conditions
\begin{equation}
    {\mathfrak{t}'}-\dot{\vartheta}=\vartheta'-\dot{\mathfrak{t}}=0.
    \end{equation}
    As discussed in \cite{Fontanella:2021btt}, these two constraints are interesting since they dictate both $\mathfrak{t}$ and $\vartheta$ satisfy a wave equation. Using the residual symmetries on the worldsheet, one can then choose a gauge where $\mathfrak{t}$ is just proportional to $\tau$, akin to a static gauge. We will next choose such an ansatz for a specific class of strings and solve the associated dynamical system. 
 \subsection{Warm up exercise: GKP string} 
 After fixing the conformal gauge, we now proceed to analyze the dynamics of a classical string configuration, choosing a periodic GKP-like \cite{Gubser:2002tv} ansatz on the sphere \cite{Floratos:2013cia}. GKP string is a folded closed string solution that has been crucial in understanding integrable features of gauge/gravity duality, and has been discussed in many integrable target space settings \footnote{Such configurations are dual to twist-two operators in the boundary theory. In the relativistic case in the usual large N ‘t Hooft limit, and the large $AdS$ spin limit, the anomalous dimension of such operators grow logarithmically with spin. This ties in nicely to the logarithmic scaling of the proper length of the bulk string solution for large spin.}. The solution is expressed as
\begin{equation}
  \mathfrak{t}=k\tau, ~~\vartheta=\vartheta(\sigma),~~\varphi=k \omega \tau
\end{equation}
Substituting this ansatz into \eqref{Lagrangian 1}, the sNC sigma model simplifies to 
\begin{equation}
    \mathcal{L}=-\frac{T}{2}\left[\sin^2{\vartheta}(-k^2\omega^2)+\xi_1(\vartheta'-\dot{\mathfrak {t}}) \right]
    \label{GKP:Lagrangian}
\end{equation}
 from which we can directly obtain the equation of motion for the Lagrange multiplier $\xi_1$, which gives us the constraint:
\begin{equation}
    \vartheta'-k=0. 
\end{equation}
Solving this yields a linear dependence of $\vartheta$ in terms of the worldsheet coordinate $\sigma$, i.e. $\vartheta=k\sigma+\sigma_0$. 
Using this solution in the equation of motion of $\vartheta$, we next solve for the Lagrange multiplier, assuming periodic boundary conditions, to get 
\begin{equation}
    \xi_1=\frac{1}{2}k\omega^2\cos{2(k\sigma+\sigma_0)}
\end{equation}
We can now proceed to find the Energy-Spin dispersion relation between the conserved Noether charges,  which can be found out in the usual way from the action:
\begin{equation}
    E=-\int_{0}^{2\pi}\frac{\partial\mathcal{L}^{NR}_{P}}{\partial \dot{t}}d\sigma=-\frac{\kappa\omega^2T}{4}\int_{0}^{2\pi}\cos{(2\kappa\sigma+2\sigma_0)}d\sigma
\end{equation}and
\begin{equation}
    J=\int_{0}^{2\pi}\frac{\partial\mathcal{L}^{NR}_{P}}{\partial \dot{\varphi}}d\sigma= \pi\omega\kappa T- \frac{\kappa\omega T}{2}\int_{0}^{2\pi}\cos{(2\kappa\sigma+2\sigma_0)}d\sigma
\end{equation}respectively. These two expressions can be appropriately scaled to find the dispersion relation as
\begin{align}
\tilde{E}-\tilde{J}=\left|\frac{\pi T}{2}\right|\,,
 \label{GKP:dispersion}
\end{align}where, $\tilde{E}=\frac{E}{\kappa\omega^2}$ and $\tilde{J}=\frac{J}{2\kappa\omega}$. This dispersion relation can now be compared with the one for relativistic cousin of the same string configuration \cite{Gubser:2002tv}. However, one should remember that in the relativistic case, the dispersion maps to the anomalous dimension of the dual operator in the explicit large spin limit, which is not the case here. 
\medskip

For circular/folded GKP strings in a relativistic $\mathbb{R}\times S^2$ background, energy $E$ and angular momentum $J$ actually assume different nontrivial dependence on the frequency $\omega$. This causes such strings in relativistic configuration to achieve two specific limits, namely, \textit{long string} and \textit{short string} limit, depending on the values of $\omega$ and the extent of the string over the curved target manifold. These two configurations of a folded rotating string on a two-sphere are inherently different. On the contrary, our GKP strings while being embedded in the intrinsic sNC $\mathbb{R}\times S^2$, do not realize any long or short string limits as in this case the energy and angular momentum vary at most polynomially with $\omega$. Thus one cannot comment on the existence of two different regimes of operators dual to these string states in the putative holographic dual field theory.

\subsection{Spinning string}
\label{section 2.3}
Let us now generalize what we did last section to our case of interest.
To study the intrinsic dynamics of the non-relativistic circular spinning strings on the two sphere, following the relativistic version  thereof, we consider the following embedding 
\begin{equation}
    X_1+iX_2=\sin \vartheta e^{i\varphi}=r_1(\sigma)e^{i\Phi(\tau,\sigma)},~~X_3=\cos \vartheta=r_2(\sigma)\label{embedding}\end{equation}
    where the sphere constraint is given by $r_1^2+r_2^2=1$ and the corresponding ansatz for the worldsheet degrees of freedom are given by:
\begin{equation}\label{spinansatz}
    \mathfrak{t}=\kappa \tau,~\vartheta=\vartheta(\sigma),~\varphi=\omega \tau+f(\sigma).
\end{equation}
Here $\kappa \in \mathbb{Z}$ and $\omega$ are just constants as in the GKP case.
Substituting the above embedding and the corresponding ansatz in the Lagrangian (\ref{Lagrangian 1}), we obtain
\begin{equation}
    \mathcal{L}_{P}^{NR}=-\frac{T}{2}\left[{r_1}^2(f'^2-\omega^2)- \xi_1 \left(\kappa-\frac{r_1^{'}}{r_2}\right)\right].
    \label{Lagrangian 2}
\end{equation}
Note that we use here $$e_\alpha=(1,1), \bar{e}_{\alpha}=(1,-1), \tau_\alpha=(1,1) \,\,\text{and} \,\,\bar{\tau}_{\alpha}=(1,-1).$$
Immediately we notice there is no quadratic kinetic term in the Lagrangian, same as the GKP case, and the only derivative of the radial field appears in the constraint. The equation of motion dictates $f(\sigma)$ is given by
$$f^{'}(\sigma)=\frac{\upsilon}{r_1^2},~~~v = \text{constant},$$
  which on substitution in \eqref{Lagrangian 2} takes the suggestive form 
  \begin{equation}
    \mathcal{L}_{P}^{NR}=-\frac{T}{2}\left[ \frac{\upsilon^2}{{r_1}^2}-\omega^2{r_1}^2-\xi_1 \left(\kappa-\frac{r_1^{'}}{r_2}\right)\right],
    \label{Lagrangian_2}
\end{equation}
which evidently contains the usual harmonic oscillator-type and centrifugal-type potentials akin to relativistic spinning string actions but does not have any kinetic term in the order of $r_1^{'2}$ as such.
The Euler-Lagrange equation corresponding to $\xi_1$ implies the constraint
\begin{align}
    \kappa-\frac{r_1^{'}}{r_2}=0 \implies r_1(\sigma)=\sin(\kappa \sigma+ \sigma_0).
    \label{eom 1}
\end{align}
This simply means that $\vartheta$ is linear in $\sigma$ coordinate \footnote{Note that for periodic solutions $\kappa \in \mathbb{Z}$ acts as a winding number. Since $t=\kappa\tau$, the string also wounds around the time direction.}.  This solution can be compared with the usual relativistic spinning string solutions on a sphere\footnote{See appendix \eqref{appendix A} for a discussion from a Neumann-Rosochatius system point of view.} where the $r$ field, still periodic in $\sigma$, is solved in terms of a Jacobi elliptic function. 
One can see in our case $\xi_1=\xi-\bar{\xi} = g(\sigma)$ from the $t$ equation of motion. The remaining Lagrange multiplier is thus a constant in worldsheet time. The equation of motion for $r_1$ gives an equation for this Lagrange multiplier:
\begin{equation}
\frac{ g'(\sigma )}{2 \sqrt{1-r_1^2}}+  \omega ^2 r_1+\frac{ v^2}{ r_1{}^3}=0.
\end{equation}Substituting $r_1(\sigma)=\sin(\kappa \sigma+ \sigma_0)$ from the constraint and doing simple algebra, we get:
\begin{equation}
    g(\sigma)=\frac{\upsilon^2 \csc ^2\left(\kappa  \sigma +\sigma _0\right)}{\kappa  }+\frac{ \omega ^2 \cos \left(2 \left(\kappa  \sigma +\sigma _0\right)\right)}{2 \kappa }.
    \label{lagrangeMULT}
\end{equation}
In all cases we assume the worldsheet fields and the Lagrange multiplier satisfy closed string boundary conditions on $\sigma$. 
\subsection*{Virasoro constraints:} 
The general Virasoro constraints from varying the action in conformal gauge \eqref{masteraction} can be written as:
\begin{align}
  H_{\mu\nu}\dot{X}^{\mu}{X'}^{\nu}+\frac{1}{2}(\xi_{0}\tau^{1}_{\mu}{X'}^{\mu}++\xi_{1}\tau^{0}_{\mu}{X'}^{\mu})=0,\\
 H_{\mu\nu}(\dot{X}^{\mu}\dot{X}^{\nu}+{X'}^{\mu}{X'}^{\nu})+\xi_{0}\tau^{0}_{\nu}{X'}^{\nu}+\xi_{1}\tau^{1}_{\nu}{X'}^{\nu}=0.
\end{align}
Upon substituting the spinning string ansatz, it reduces to\footnote{Note that the 
transformation $\theta\rightarrow -\theta$ alone is not a symmetry of the non-relativistic action \eqref{masteraction}. However, if we simultaneously exchange the worldsheet zweibeins and Lagrange multipliers, i.e $e_{\alpha}\leftrightarrow \bar{e}_{\alpha},~\xi\leftrightarrow \bar{\xi},~\tau_{\mu}\leftrightarrow \bar{\tau}_{\mu}$ the non-relativistic action remains invariant.}
\begin{align}
   2 H_{\varphi\varphi}\dot{\varphi}\varphi'-\xi_{0}\vartheta'=0,\\
    H_{\varphi\varphi}(\dot{\varphi}^2+{\varphi'}^2)-\xi_{1}\vartheta'=0.
\end{align}
From these two constraints, one can deduce,\begin{equation}
    \xi_1= \frac{\omega^2}{\kappa}\sin^2(\kappa\sigma+\sigma_0)+\frac{v^2}{\kappa\sin^2{(\kappa\sigma+\sigma_0)}}
\end{equation}
Such expression of $\xi_1$ can be retrieved from the expressions of $g(\sigma)$ in (\ref{lagrangeMULT}) by appropriately choosing the integration constant. The other Lagrange multiplier $\xi_0$ can be obtained from the Virasoro constraint as $\xi_0=\frac{2\omega v}{\kappa}$. Thus, $\xi_0=0$ when we choose $v=0$, i.e. we go back to the GKP case.

\subsection*{Conserved charges and dispersion relation}
The conserved energy and angular momentum can be derived respectively as the Noether charges corresponding to the cyclic coordinates $t$ and $\varphi$, with $d\sigma = d\vartheta/\vartheta'$:
\begin{subequations}
    \begin{align}
        &E=-\int_{0}^{2\pi}\frac{\partial\mathcal{L}^{NR}_{P}}{\partial \dot{t}}d\sigma=-\frac{Tv^2}{2\kappa^2}\int_{\vartheta_{min}}^{\vartheta_{max}}\frac{d\vartheta}{\sin^2{\vartheta}}+\frac{\omega^2}{2\kappa^2}\int_{\vartheta_{min}}^{\vartheta_{max}}\cos(2\vartheta)d\vartheta\\&
        J=\int_{0}^{2\pi}\frac{\partial\mathcal{L}^{NR}_{P}}{\partial \dot{\varphi}}d\sigma=\pi T
-\frac{T}{2}\int_{\vartheta_{min}}^{\vartheta_{max}} \cos(2\vartheta)d\vartheta   \end{align}
\end{subequations}
However, there are more elementary string theory excitations than closed folded strings on the two-sphere, like the \textit{Giant Magnons} put forward by Hofman and Maldacena \cite{Hofman:2006xt}, which are bulk duals of magnon excitations in the sYM \footnote{In the spin chain description of $\mathcal{N}=4$ sYM, magnons appear as excitations or impurities on top of the ferromagnetic vacuum of $SU(2)$ sector \cite{Beisert:2005tm}. The bulk dual object is that of a string soliton wrapping the equator of $S^2$ but travelling with a finite momentum.}.
The dual magnon momenta for such configurations are given by conserved deficit angle $\Delta\varphi$ associated to the extent of the angular coordinate, and can be evaluated as 
\begin{equation}
    \Delta\varphi=\int f'(\sigma)d\sigma=\frac{v}{\kappa^2}\int_{\vartheta_{min}}^{\vartheta_{max}}\frac{d\vartheta}{\sin^2{\vartheta}}
\end{equation}
In usual relativistic case, magnons move between some restricted angular extent on the sphere, however in sNC case nothing stops it from reaching the poles from the equator.
Looking at these expressions, one gets the energy-angular momentum dispersion relation 
\begin{equation}
    \frac{2\kappa}{\omega^2}E-J=\frac{v}{\omega^2}T\Delta \varphi.
\end{equation}
With appropriate scaling, this becomes,
\begin{equation}
    \hat{E}-\hat{J}\sim T\Delta\hat\varphi\,.
\end{equation}
Note that this is obviously different from the usual (strong coupling) Giant Magnon dispersion relation on the relativistic two sphere: $E-J = T|\sin\frac{\Delta\varphi}{2}|,~p_{magnon}\sim\Delta\varphi$, but can be thought of a small momentum ($p_{magnon}\ll 1$) limit of the same.
This is somewhat similar to the case for the scaling obtained with spinning string in the tNC formalism over $\mathbb{R}\times S^2$ \cite{Roychowdhury:2020kma}, where the non-relativistic magnon dispersion has been shown as a $c\to \infty$ limit of the relativistic case\footnote{In this sense the interpretation as a Giant Magnon solution is not entirely clear, since as the worldsheet momentum is vanishing,
the soliton (the kink) should shrink to zero size, and reduce to a massless point particle moving around on the equator of the sphere with the angular momentum equal to $J$.}. 

\subsection*{Other string configurations}
Now let us talk about more general solutions for non-relativistic sigma model. More specifically, the embedding of a Giant Magnon like solitonic solution requires one to add a characteristic velocity of the excitation or a wave propagating along the string. 
Let us choose such a more generic solitonic ansatz as 
\begin{equation}
   \mathfrak{t}=\kappa\tau,~ \vartheta=\vartheta(y),~\varphi=\omega\tau+f(y),~ y=\sigma-\beta\tau, 
   \label{general ansatz}
\end{equation}
where the spinning string embedding assumes a finite ``velocity'' parameter $\beta$. However, the equations of motion for auxiliary fields $\xi_0$ and $\xi_1$ from the sNC Lagrangian (\ref{Lagrangian 1}) gives for this ansatz:
\begin{equation}
    \beta\vartheta'(y)=\vartheta'(y)-\kappa=0
\end{equation}
This subsequently gives $\beta=0$ which reduces the ansatz (\ref{general ansatz}) into that previously taken in (\ref{spinansatz}) and hence the overall sNC framework reproduces exactly same spinning solutions as we obtained in section \eqref{section 2.3}. Therefore it is evident that the sNC structure of the $\mathbb{R}\times S^2$ only accommodates static solitonic ansatz with $\beta=0$ for the probe string.
\medskip

To round-up our discussions of fundamental string solutions, one could also observe the following string ansatz, a $\tau \leftrightarrow \sigma$ interchanged version of \eqref{spinansatz} is also a viable one:
\begin{equation}
    X_1+iX_2=\sin \vartheta e^{i\varphi}=r_1(\tau)e^{i\Phi(\tau,\sigma)},~~X_3=\cos \vartheta=r_2(\tau)\label{embed:pulsating}
\end{equation}
with the embedding functions, $\mathfrak{t}=\tilde{\kappa} \tau,~\vartheta=\vartheta(\tau),~\varphi=m\sigma+\psi(\tau)$,
where $m$ represents the winding number along the azimuthal direction. Note that with such an ansatz one would also be able to in principle keep both Lagrange multipliers in the sNC dynamics of \eqref{masteraction}.  
Hence the phase space associated to classical sNC strings on the two-sphere itself is pretty large, despite the non-compatibility of certain embeddings borrowed from relativistic intuition. To understand more about these classical solutions, we will adopt a separate approach, based on the string $1/c^2$ expansion \cite{Hartong:2021ekg} of the sigma model action in the next section.
\section{Strings in sNC target space from $\frac{1}{c^2}$ expansion }
\label{section 3}
In this section, to study non-relativistic string theory we employ the systematic string $1/c^2$ expansion of the relativistic string introduced in \cite{Hartong:2021ekg, Hartong:2022dsx}, where $c$ acts as the speed of light in target space. This expansion singles out two longitudinal directions to be scaled accordingly, and then taking the large $c$ limit leads to the sNC geometry admitting a 2-dimensional Lorentzian submanifold\footnote{This $c$-expansion can be thought of an effective expansion, where one can devise a dimensionless parameter involving a compact length scale \cite{Hartong:2021ekg}. }. Once the geometric data is plugged back in the string action, the action itself can be expanded in orders of $1/c^2$.  Before proceeding, we first quickly review the expansion method developed in \cite{Hartong:2021ekg}.
\medskip

Although \cite{Hartong:2021ekg,Hartong:2022dsx} discuss both Nambu-Goto and Polyakov formalisms, our starting point is  the Polyakov sigma model action for a relativistic fundamental string, with all relevant $c$ factors restored:
\begin{equation}
    S=-\frac{cT}{2}\int d\tau d\sigma \left[\sqrt{-h}
h^{ab}\mathcal{G}_{MN}\partial_a X^M\partial_b X^N \right],
\end{equation}
where $\mathcal{G}_{MN}(X)$ is the induced metric of the background where the string moves.
Along with the embedding pullback fields, we are considering the worldsheet expansions of the form in leading order (LO) and next-to-leading order (NLO):
\begin{align}
X^{M}=x^{(0)}+c^{-2}x^{(1)}+\mathcal{O}\left(c^{-4}\right),\quad
   h_{ab}= h^{(0)}_{ab}+c^{-2}h^{(1)}_{ab}+\mathcal{O}\left(c^{-4}\right).  
\end{align}
We also keep ourselves confined to even power expansion in $1/c$ following the structure laid out in \cite{Hartong:2021ekg}.
With this, the pullback metric takes the following expansion
\begin{equation}
    G_{ab}=c^2\tau_{ab}+H_{ab}+\mathcal{O}(c^{-2})
\end{equation}
Note these two longitudinal and transverse boost-invariant metrics $\tau_{ab}$ and $H_{ab}$ appears in the intrinsic formulation of section \eqref{section 2} naturally. 
We have further gauge fixed the LO and NLO worldsheet by setting ${h^{(0)}_{ab}}=\eta_{ab}$ with determinant $\sqrt{-h^{(0)}}=e$\footnote{Note that this is only a partial gauge fixing since Lorentzian LO worldsheet metric behaves similarly as that of relativistic Polyakov action. In the expanded actions of our interest, only ${h^{(0)}_{ab}}$ appears in the measure.}.
The Polyakov Lagrangian then takes the expanded form,
$$ \mathcal{L}=c^2\mathcal{L}_{LO}+\mathcal{L}_{NLO}+c^{-2}\mathcal{L}_{NNLO}+\mathcal{O}\left(c^{-4}\right)
  \label{lagrangian},$$ where
\begin{align}
    \mathcal{L}_{LO}&=-\frac{eT_{eff}}{2}h^{(0)ab}\tau_{ab},\\
    \mathcal{L}_{NLO}&=-\frac{eT_{eff}}{2}h^{(0)ab}H_{ab}+\frac{eT_{eff}}{4}\mathcal{G}^{(0)abcd}\tau_{ab}h^{(1)}_{cd}
    +x^{(1)}\frac{\delta \mathcal{L}_{LO}}{\delta x^{(0)}},\label{nloexp}
\end{align}
given the Wheeler-DeWitt metric is given by
 $\mathcal{G}^{(0)abcd}=h^{(0)ac}h^{(0)bd}+h^{(0)ad}h^{(0)bc}-h^{(0)ab}h^{(0)cd}$ and we use a rescaled tension $T_{eff}=cT$. At NLO order, both $h^{(0)}$ and $h^{(1)}$ are present and varying with the latter  produces the LO Virasoro constraint,
while varying with the former leads to the NLO Virasoro constraint. For more details on constraint structures, the reader is requested to consult the original set of papers. 
To be in tune with the previous section, we choose the target space as the simplest $\mathbb{R}\times S^2$ which assumes the Lorentzian metric
 \begin{equation}
    ds^2=-dt^2+ d\theta^2+\sin^2{\theta}d\phi^2.
    \label{metric2}
 \end{equation} 
 
Following a similar approach to the one described above, we carry out the $1/c^2$ expansion for strings in this target space, considering the expansion of the pullback coordinates as follows:
\begin{align}
t&=t^{(0)}+{c^{-2}}t^{(1)}+\mathcal{O}\left({c^{-4}}\right)\label{t},\\
\theta&=\theta^{(0)}+{c^{-2}}\theta^{(1)}+\mathcal{O}\left({c^{-4}}\right)\label{theta},\\
\phi &= \phi^{(0)}+{c^{-2}}\phi^{(1)}+\mathcal{O}\left({c^{-4}}\right)\label{phi}.
\end{align}

Putting this back into the action, LO and NLO Polyakov Lagrangian finally can be written as
\begin{align}
    \mathcal{L}_{LO}&=\frac{eT_{eff}}{2}\left[(t^{(0)})'^2-(\dot{t}^{(0)})^2-(\theta^{(0)})'^{2}+(\dot{\theta^{(0)}})^2\right],\label{LOLag}\\
    \mathcal{L}_{NLO}&=\frac{eT_{eff}}{2}\bigg[2(t^{(0)})'(t^{(1)})'-2(\theta^{(0)})'(\theta^{(1)})'-2(\dot{t}^{(0)})(\dot{t}^{(1)})+2(\dot{\theta^{(0)}})(\dot{\theta^{(1)}})\nonumber\\ & -\sin^2{\theta^{(0)}}((\phi^{(0)})'^{2}-(\dot{\phi^{(0)}})^2)  \bigg].
    \label{expLagr}
\end{align}
Note that we have, without loss of generality, used a further local gauge fixing $h^{(1)}_{ab} = 0$ to write these actions. One can easily continue expanding these actions to NNLO and beyond, but the above will be enough for our discourse.
In what follows we will discuss particular string solutions for these two actions. Connections with the intrinsic sNC formalism will be discussed as need arises.  
\section{ Spinning string in $\frac{1}{c^2}$-expanded sNC target space}
\label{section 4}
Once we have our expanded actions, we can think of specific solutions embedded order by order. It should be intuitively clear to us that the non-relativistic setting is not really compliant with most non-trivial embeddings. Simplest examples of such solutions includes the well-known BMN-string solution \cite{Berenstein:2002jq}, where the target space time and angular coordinates are taken simply proportional to worldsheet time. A generalization is the so called spinning string of section \eqref{section 2} where, in the most general case, one takes the embedding to be generic functions of worldsheet coordinates: 
\begin{equation}
    t=t(\tau,\sigma),\quad \theta=\theta(\sigma), \quad \phi=\phi(\tau,\sigma)
\end{equation}
However, now we are dealing with such strings at every order of the expanded action, so we need to be careful about the ansatz. We can understand this a bit better by substituting (\ref{t})-(\ref{phi}) which generates the following structure for leading and subleading string ansatz:
\begin{equation}
    t^{(0)}= \kappa_0 \tau,~~ t^{(1)}= \kappa_{1}(\sigma,\tau) ,~~\theta^{(i)}=\theta^{(i)}(\sigma),~~\phi^{(i)}=\phi^{(i)}(\tau,\sigma),\quad i=0,1.
    \label{ansatzsp}
\end{equation}
Note that the leading order embedding for $t$ is just that of static gauge with $\kappa_0$ being a constant. 
By substituting the worldsheet dependencies of the string ansatz, we obtain the final expression for the LO and NLO Lagrangian as
\begin{align}
    \mathcal{L}_{LO}&=-\frac{e T_{eff}}{2}\left[(\dot{t}^{(0)})^2
 +({\theta^{(0)}}')^2\right]\label{LOA},\\
   \mathcal{L}_{NLO}&=-\frac{e T_{eff}}{2}\Bigg[2\dot{t}^{(0)}\dot{t}^{(1)} + 2(\theta^{(0)})'(\theta^{(1)})'+\sin^2{\theta^{(0)}}\Big(({\phi^{(0)}}')^2-(\dot{\phi}^{(0)})^2\Big)\Bigg].
   \label{NLO}
   \end{align}
From our idea of relativistic spinning strings in \eqref{metric2}, a suitable embedding for the pullback functions in the radial and azimuthal directions (in parent target space) can be taken as
\begin{equation}
  X_1+iX_2=\sin{\theta} e^{i\phi}=r_1(\sigma)e^{i\Phi(\tau, \sigma)},~~X_3=\cos{\theta}=r_2(\sigma),~~\text{and} ~~\Phi(\tau,\sigma)=f(\sigma)+\omega\tau.
\label{embed2}\end{equation}
In the relativistic case these $r_i$'s are related by the sphere constraint. To chalk out the constraints associated to expanded actions, we also need to consider the $1/c^2$ expansion of these embedding coordinates in a similar vein as those of pullback fields, which reads:
\begin{equation}
    r_1(\sigma)=r^{(0)}_1(\sigma)+c^{-2}r^{(1)}_1(\sigma)+\mathcal{O}(c^{-4}),~~r_2(\sigma)=r^{(0)}_2(\sigma)+c^{-2}r^{(1)}_2(\sigma)+\mathcal{O}(c^{-4})
    \label{rexp}
\end{equation}
Now comparing \eqref{embed2} and \eqref{rexp}, we have the relevant fields at different orders:
\begin{align}
r^{(0)}_1(\sigma)=\sin{\theta^{(0)}},~~r^{(1)}_1(\sigma)=\cos{\theta^{(0)}}\theta^{(1)},~~r^{(0)}_2(\sigma)=\cos{\theta^{(0)}},~~r^{(1)}_2(\sigma)=-\sin{\theta^{(0)}}\theta^{(1)}
   \label{embed}
\end{align}
Note that the leading order constraint with $\theta^{(0)}=\sin^{-1}(r^{(0)}_1(\sigma))$ remains that of the original sphere, while subleading terms lead to a new constraint: 
$$(\theta^{(1)})^2=(r^{(1)}_1(\sigma))^2+(r^{(1)}_2(\sigma))^2$$ Let us, for completeness, mention all other constraint relations in both leading and next-leading embedding coordinates:
\begin{equation}
\begin{split}
(r_1^{(0)})^2+(r_2^{(0)})^2=1,~~r_1^{(0)} (r_1^{(0)})'+r_2^{(0)} (r_2^{(0)})' =0 ,\\
      r_1^{(0)}r_1^{(1)}+r_2^{(0)}r_2^{(1)}=0,\quad r_1^{(1)} (r_1^{(0)})'+r_1^{(0)} (r_1^{(1)})'+r_2^{(1)} (r_2^{(0)})'+r_2^{(0)} (r_2^{(1)})'=0.
      \label{Constraints}
      \end{split}
  \end{equation}
  Clearly not all of the above constraints are independent, but they give us an idea how the dynamical systems on leading and subleading spheres are related to each other. These will be important for us in what follows.
\subsection{LO dynamics}
\label{subsection 4.1}
With our setup in place, let us discuss string solutions at LO and NLO orders of the expanded action. In the leading order, the action is only dependent on the longitudinal directions \eqref{LOA}.
A quick check using the equations of motion confirms that the leading order ansatz derived from \eqref{embed2} and supplemented by $t^{(0)}= \kappa_0 \tau$ is a consistent one.
Substituting the ansatz, our action becomes
\begin{eqnarray}
\begin{split}
    \mathcal{L}_{LO}=
    -\frac{e T_{eff}}{2}\left[\kappa_0^2+\frac{{r'^{(0)}_1(\sigma)}^2}{{1-r^{(0)}_1(\sigma)}^2} \right]
    \label{LO Lagrangian}
    \end{split}
\end{eqnarray}
The conserved quantities are given by,
\begin{equation}
  E_{0}=e T_{eff}\kappa_0  ,\quad \pi_{r^{(0)}_1}=-e T_{eff}\frac{r'^{(0)}_1(\sigma)}{{1-r^{(0)}_1(\sigma)}^2} .
    \label{pir0}
\end{equation}
The canonical Hamiltonian is given by,
\begin{equation}
    \begin{split}
    \mathcal{H}_{LO}=(r_1^{(0)})^{'}\pi_{r^{(0)}_1}-\mathcal{L}_{LO}=\frac{1}{2}\left[e T_{eff}\kappa_0^2-\frac{1}{eT_{eff}}(\pi_{r_1^{(0)}})^2(1-(r_1^{(0)})^2)\right].
    \end{split}
    \label{LO Hamiltonian}
\end{equation}
In a straightforward calculation, Euler-Lagrange equation of motion for $r^{(0)}_1$ from the Lagrangian $\mathcal{L}_{LO}$ yields,
\begin{equation}
    r^{(0)}_1=\sin \left(a\sigma+b\right),~~r^{(0)}_2=\cos \left(a\sigma+b\right)
    \label{LOSOL}
\end{equation}where we used the condition for closed string as $r^{(0)}_i(\sigma=0)=r^{(0)}_i(\sigma=2\pi)$ with $a \in \mathbb{Z}$ acting as a winding number around $\theta$ direction. This is clearly a set of oscillating periodic solutions. Further, one can compare this solutions with the ones we got from the Lagrange multiplier in the intrinsic sNC case of \eqref{eom 1}. This should be more clear going forward.
\medskip

Let us pause and comment on few things here. In usual spherical spinning strings one gets oscillating solutions in terms of the Jacobi elliptic functions, so we may also notice, our LO solutions can be written as:
\begin{equation}
    r^{(0)}_1=
\text{sn}~(u|k),~~r^{(0)}_2=\text{cn}~(u|k),~~a\sigma+b=\text{Amplitude of u}
\end{equation}Thus, by definition, we can write $u$ in the standard form of the integration as
\begin{equation}
    u=\int_0^{(a\sigma+b)}\frac{d\alpha}{\sqrt{1-k^2\sin^2\alpha}}
    \label{u}
\end{equation} 
 For relativistic solutions to the embeddings of spinning string over $\mathbb{R}\times S^2$, albeit from a Neumann-Rosochatius system point of view, one can see Appendix \eqref{appendix A}. Note that by definition $
\text{sn}(u|0)=\sin(u)$  and $
\text{cn}(u|0)=\cos(u)$, which helps one to extract the LO solutions by using the series expansion of the Jacobi sn function around small values of $k$:
\begin{equation}
    \text{sn}(u|k)=\sin{u}+\frac{1}{8}\cos{u}\left(-2u+\sin{2u}\right)k+\mathcal{O}(k^2)
\end{equation}
In large $c$ limit, the relativistic radial coordinate $r_1$ can be expanded as (\ref{rexp}). Therefore, the series expansion with respect to $k\to 0$ of the relativistic Jacobi sn solution can be mapped to the solutions in non-relativistic limit, at least up to the leading order. Note further that only the LO radial dynamics can be formulated in the same exactly solvable dynamical system language which is trivially integrable, and more discussions can be found in Appendix \eqref{appendix C}.
\subsection{NLO dynamics and Neumann-Rosochatius-like model}
\label{subsection 4.2}
\begin{center}
    \begin{figure}[!]
\includegraphics[scale=0.38]{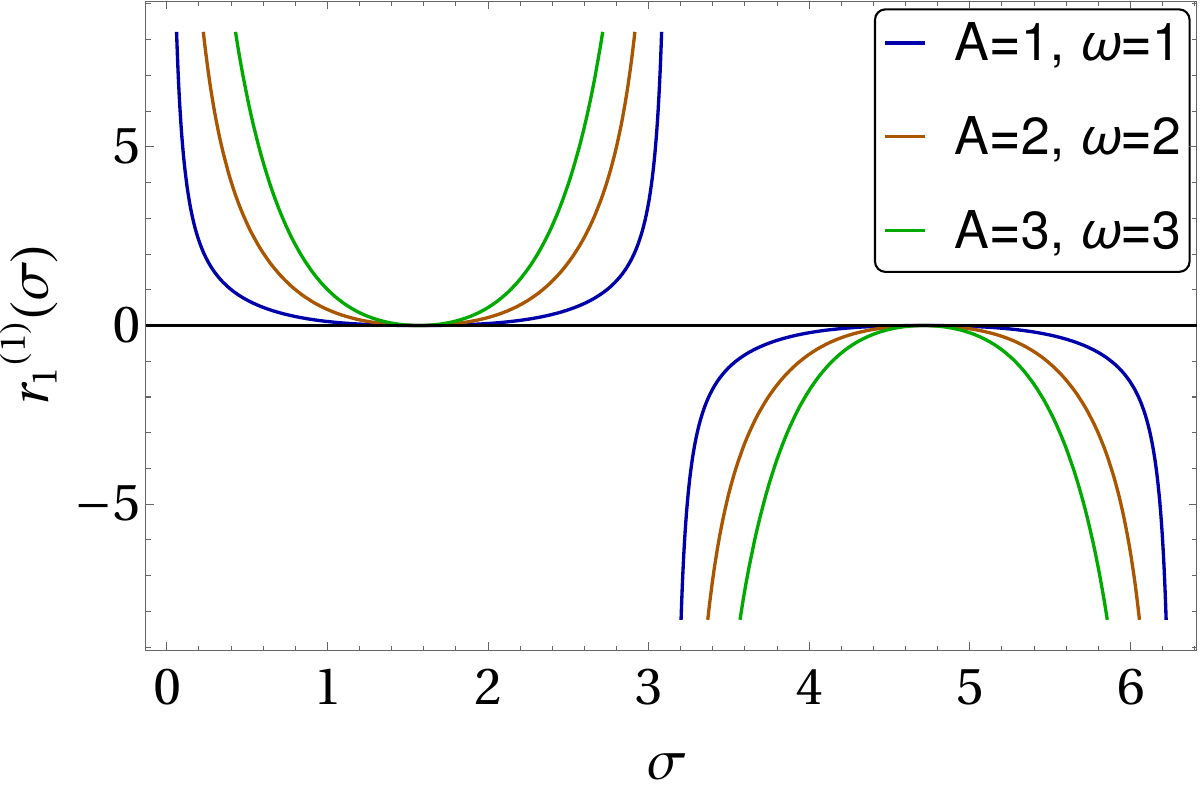}
\hfill
\includegraphics[scale=0.38]{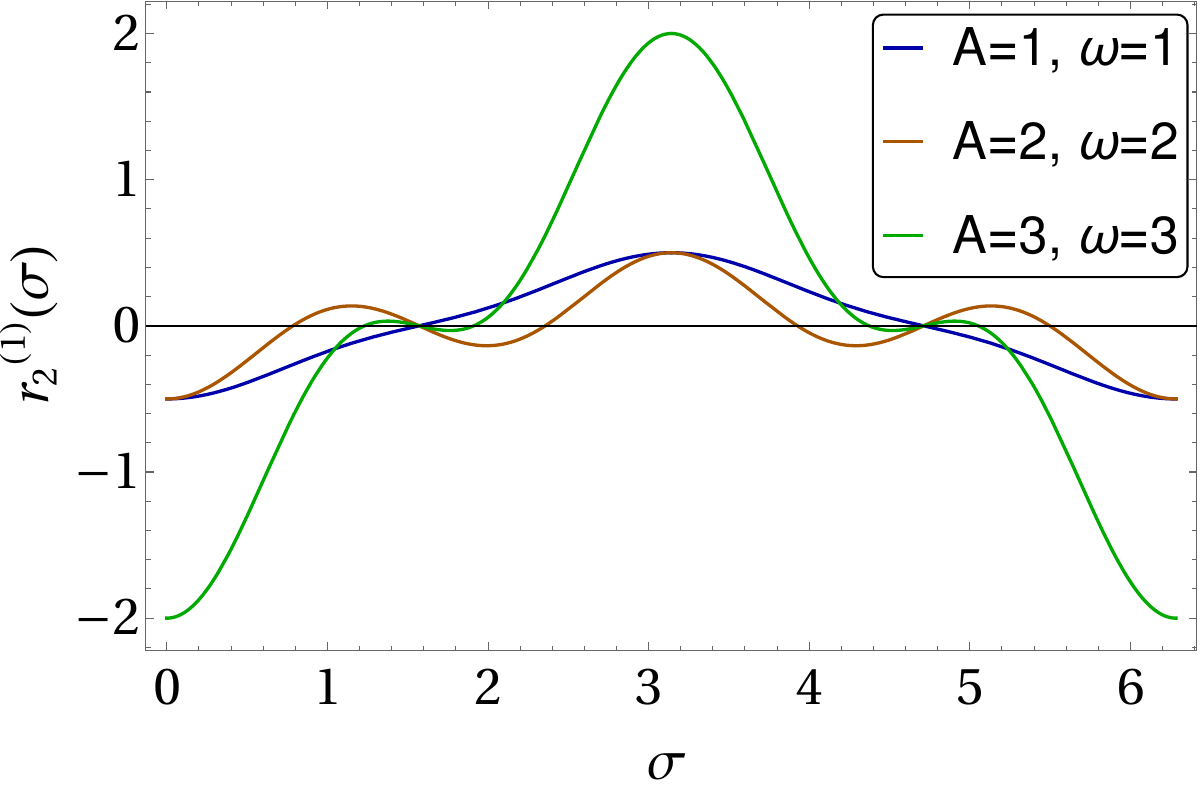}
\caption{$r_1^{(1)}(\sigma)$(left) and $r_2^{(1)}(\sigma)$(right) for the spinning string solution. Here $a=1$ and $b=\pi$, and we use different values of the other parameters.}
\label{fig:1}
\end{figure}
\end{center}

At the NLO the Lagrangian is given by \eqref{NLO}, for which to solve the EOM we need to explicitly put back the leading order solutions for the fields discussed in the last section.  Doing so, a little algebra gives us:
\begin{equation}
\begin{split}
    &
    t^{(1)}=h(\sigma)\tau,\quad
    f'(\sigma)=\frac{A}{\sin^2{\theta^{(0)}}}\implies 
    f(\sigma)=\frac{A \cot (a\sigma+b  )}{a},\quad
    \end{split}
    \end{equation}
    where $h(\sigma)$ is an undetermined function at this level. 
Further using the relations \eqref{Constraints} we can transform the NLO action to an action for embedding coordinates, involving both LO and NLO radial fields:
\begin{equation}
    \mathcal{L}_{NLO}=-\frac{eT_{eff}}{2}\Bigg[2 \dot{t}^{(0)}\dot{t}^{(1)}+2\frac{(r_1^{(0)})'}{\sqrt{1-(r_1^{(0)})^2}}\left( \frac{r_1^{(1)}(r_1^{(1)})'+r_2^{(1)}(r_2^{(1)})'}{\sqrt{(r_1^{(1)})^2+(r_2^{(1)})^2}}\right)+\frac{A^2}{(r_1^{(0)})^2}-\omega^2(r_1^{(0)})^2 \Bigg].
    \label{nloL}
\end{equation}
Now, we want to focus on the radial dynamics, and the EOM for $r_1^{(0)}$ gives us
\begin{equation}
    \frac{d}{d\sigma}\left(\frac{r_1^{(1)}(r_1^{(1)})'+r_2^{(1)}(r_2^{(1)})'}{\sqrt{(r_1^{(1)})^2+(r_2^{(1)})^2}}\right)=-\left(\frac{A^2}{(r_1^{(0)})^3}+\omega^2r_1^{(0)}\right)r_1'^{(0)}\,,
   \label{r10 equation}
\end{equation}
which readily leads to:
\begin{equation}
    \left(\sqrt{(r_1^{(1)})^2+(r_2^{(1)})^2}\right)'=\pm\theta_1'(\sigma)=-\left[\frac{\omega^2}{4a}\cos{(2(a\sigma+b))}+\frac{A^2}{2a}\frac{1}{\sin^2{(a\sigma+b)}}\right]
\end{equation}
Here, choosing the relevant sign, we observe that $\theta_1'$ exhibits a behaviour similar to that of the Lagrange multiplier $\xi_1$ calculated in \eqref{lagrangeMULT}. This requires a comparison between the parameters as: $a=\frac{\kappa}{2},~A={v}$.
Thus, integrating for $\theta_1$, we can easily get the solutions for the NLO fields $r_1^{(1)}$ and $r_2^{(1)}$ using the relevant constraints, yielding 
\begin{equation}
    r_1^{(1)}=\cos (a\sigma+b)~~\theta^{(1)}(\sigma)=\cos (a\sigma+b)\left[\frac{-4 A^2 \cot (a\sigma+b)+{\omega}^2 \sin (2(a\sigma+b))}{8 a^2}\right],
    \label{r11 eqn}
\end{equation}and
\begin{equation}
    r_2^{(1)}=-\sin{(a\sigma+b)}~\theta^{(1)}(\sigma)=-\sin{(a\sigma+b)}\left[\frac{-4 A^2 \cot (a\sigma+b)+{\omega}^2 \sin (2(a\sigma+b))}{8 a^2}\right].
    \label{r21eqn}
\end{equation}
We plot these NLO fluctuations of the radial coordinates assuming again the periodicity of $\sigma $ for closed string. The periodic behaviours of these fluctuations are shown in Figure \eqref{fig:1} for different parameter values. 
\medskip
Before moving on, let us comment on the connection between the intrinsic sNC action and the NLO expanded action. It has been shown in \cite{Hartong:2022dsx} that the intrinsic sNC string can be mapped to the Polyakov NLO expanded one, provided the term proportional to $x^{(1)}$ identically vanishes in \eqref{nloexp}\footnote{This may be a bit too quick an argument. In fact, there are general cases where this holds true. But for the current case, the identification is easily established.}. This proves the equivalence of the NLO action with the Gomis-Ooguri action \cite{Gomis:2000bd} as well. In our discussion, we have clearly seen that the intrinsic sNC action contains the LO solution as dynamical constraints, and the information of NLO fields are embedded in the Lagrange multiplier. Said in other words, NLO fields, which appear with single derivatives in \eqref{expLagr}, can only be solved when we put the LO ones \textit{on-shell}. 
\medskip

It is worth noting that, similarly to an integrable Neumann-Rosochatius model on a sphere, the Lagrangian (\ref{nloL}) obtained for the NLO dynamics includes both the $r^2$ and $\frac{1}{r^2}$-type potentials (see Appendix \eqref{appendix A}), although the kinetic term is more intricate. 
Thus, one could work out that the NLO dynamics of spinning string in $\mathbb{R}\times S^2$ can be described equivalently by a similar exactly solvable model which is a deformation of the classical integrable Neumann-Rosochatius model. The radial coordinates $r_1^{(0)}$ (or $r_2^{(0)}$)  and $\tilde{r}=\sqrt{(r_1^{(1)})^2+(r_2^{(1)})^2}$ of such a model equivalently describes string dynamics on the LO and NLO spheres with $\theta^{(0)}$ and $\theta^{(1)}$ respectively. This makes even more sense when we remind ourselves that the analysis of diffeomorphism in conformal gauge for both the LO and NLO orders shows that the algebra generated by residual gauge transformations will always be two copies of the Witt algebra. This makes sure the canonical worldsheet Hamiltonian always vanishes, and the conformal constraint gives rise to an exact radial Hamiltonian. We can now attempt to write down this equivalent NLO radial dynamical system.
\medskip 

The canonical momenta $\pi_{r_1^{(0)}}$, $\pi_{r_1^{(1)}}$, $\pi_{r_2^{(1)}}$ and $\pi_{t_1^{(1)}}$ for $r_1^{(0)}$, $r_1^{(1)}$, $r_2^{(1)}$ and $t_1^{(1)}$ are respectively derived from the Lagrangian (\ref{nloL}):
\begin{subequations}
\begin{align}
   &\pi_{r_1^{(0)}}=-eT_{eff}\left(\frac{r_1^{(1)}r_1'^{(1)}+r_2^{(1)}r_2'^{(1)}}{r_2^{(0)}\sqrt{(r_1^{(1)})^{2}+(r_2^{(1)})^{2}}}\right),\nonumber\\& 
   \pi_{r_1^{(1)}}=-eT_{eff}\left(\frac{r_1^{(1)}r_1'^{(0)}}{r_2^{(0)}\sqrt{(r_1^{(1)})^{2}+(r_2^{(1)})^{2}}}\right)=-eT_{eff}r_1'^{(0)},\nonumber\\&
   \pi_{r_2^{(1)}}=-eT_{eff}\left(\frac{r_2^{(1)}r_1'^{(0)}}{r_2^{(0)}\sqrt{(r_1^{(1)})^{2}+(r_2^{(1)})^{2}}}\right)=eT_{eff}\frac{r_1^{(0)}r_1'^{(0)}}{r_2^{(0)}},\nonumber\\&
   \pi_{t_1^{(1)}}=-eT_{eff}\kappa
\end{align}
\label{canonical momenta for spinning}
\end{subequations}
However, this is an extremely constrained system due to the conditions between fields of different orders in \eqref{Constraints}. 
In fact the NLO canonical momenta $\pi_{r_1^{(1)}}$ and $\pi_{r_2^{(1)}}$, on using $r_1^{(0)}r_1^{(1)}=-r_2^{(0)}r_2^{(1)}$,  follow the additional relation
\begin{equation}
    \pi_{r_1^{(1)}}r_2^{(1)}+\pi_{r_2^{(1)}}r_1^{(1)}=0.
    \label{momenta rel}
\end{equation}
It is also crucial to note that $\pi_{r_1^{(0)}}=-eT_{eff}\left(r_1'^{({1})}-\frac{r_1^{(0)}}{r_2^{(0)}}r_2'^{(1)}\right)$, i.e., it contains both $r_1'^{(1)}$ and $r_2'^{(1)}$.
One can find the above equations by starting with an equivalent Neumann-Rosochatius-like solvable system whose Hamiltonian assumes the form 
\begin{equation}
    H_{NR}= -\frac{1}{eT_{eff}}\left(\pi_{r_1^{(0)}}^{(1)}\pi_{r_1^{(1)}}-\frac{r_2^{(0)}}{r_1^{(0)}}\pi_{r_1^{(0)}}^{(2)}\pi_{r_2^{(1)}}\right)+\frac{eT_{eff}}{2}\Bigg[\frac{A^2}{(r_1^{(0)})^2}-\omega^2(r_1^{(0)})^2 \Bigg]
    \label{NRHnew}
\end{equation}where, $\pi_{r_1^{(0)}}=\pi_{r_1^{(0)}}^{(1)}+\pi_{r_1^{(0)}}^{(2)}$. 
From this Hamiltonian, the equations $\frac{\partial H_{NR}}{\partial \pi_{r_1^{(1)}}}=r_1'^{(1)}$ and $\frac{\partial H_{NR}}{\partial \pi_{r_2^{(1)}}}=r_2'^{(1)}$ yield 
\begin{equation}
    \pi_{r_1^{(0)}}^{(1)}=-eT_{eff}r_1'^{(1)},~~\pi_{r_1^{(0)}}^{(2)}=eT_{eff}\frac{r_1^{(0)}}{r_2^{(0)}}r_2'^{(1)}
    \label{nlopi's}
\end{equation}
Therefore we consistently get the expression of $\pi_{r_1^{(0)}}$ by adding these two expressions in (\ref{nlopi's}).
Note that provided the constraints between fields and momenta, the above system can be equivalently written as
\begin{equation}
    H_{NR}= - \frac{1}{eT_{eff}}\pi_{r_1^{(0)}}\pi_{r_1^{(1)}}+\frac{eT_{eff}}{2}\Bigg[\frac{A^2}{(r_1^{(0)})^2}-\omega^2(r_1^{(0)})^2 \Bigg]
     \label{NRH}
\end{equation}

The corresponding Hamilton's equation of motion $\frac{\partial H_{NR}}{\partial\pi_{r_1^{(0)}}}=r_1'^{(0)}$ gives $\pi_{r_1^{(1)}}$ and $\frac{\partial H_{NR}}{\partial r_1^{(0)}}=-\pi'_{r_1^{(0)}}$ exactly reproduces \eqref{r10 equation}. We can similarly get the expression of $\pi_{r_2^{(1)}}$ by using the constraints $r_1^{(0)}r_1^{(1)}+r_2^{(0)}r_2^{(1)}=0$ and $\pi_{r_1^{(1)}}r_2^{(1)}+\pi_{r_2^{(1)}}r_1^{(1)}=0$.  One could thus claim that \eqref{NRHnew} gives a solvable system for strings on the NLO sphere. However, note that due to the constraints, the structure of integrals of motion in this case is quite nontrivial. Some comments regarding both LO and NLO integrals of motion can be found in Appendix \eqref{appendix C}.

\subsection{Bohr-Sommerfeld quantization}
\label{subsection 4.3}
\begin{figure}[h!]
     \centering
     \includegraphics[width=0.41\linewidth]{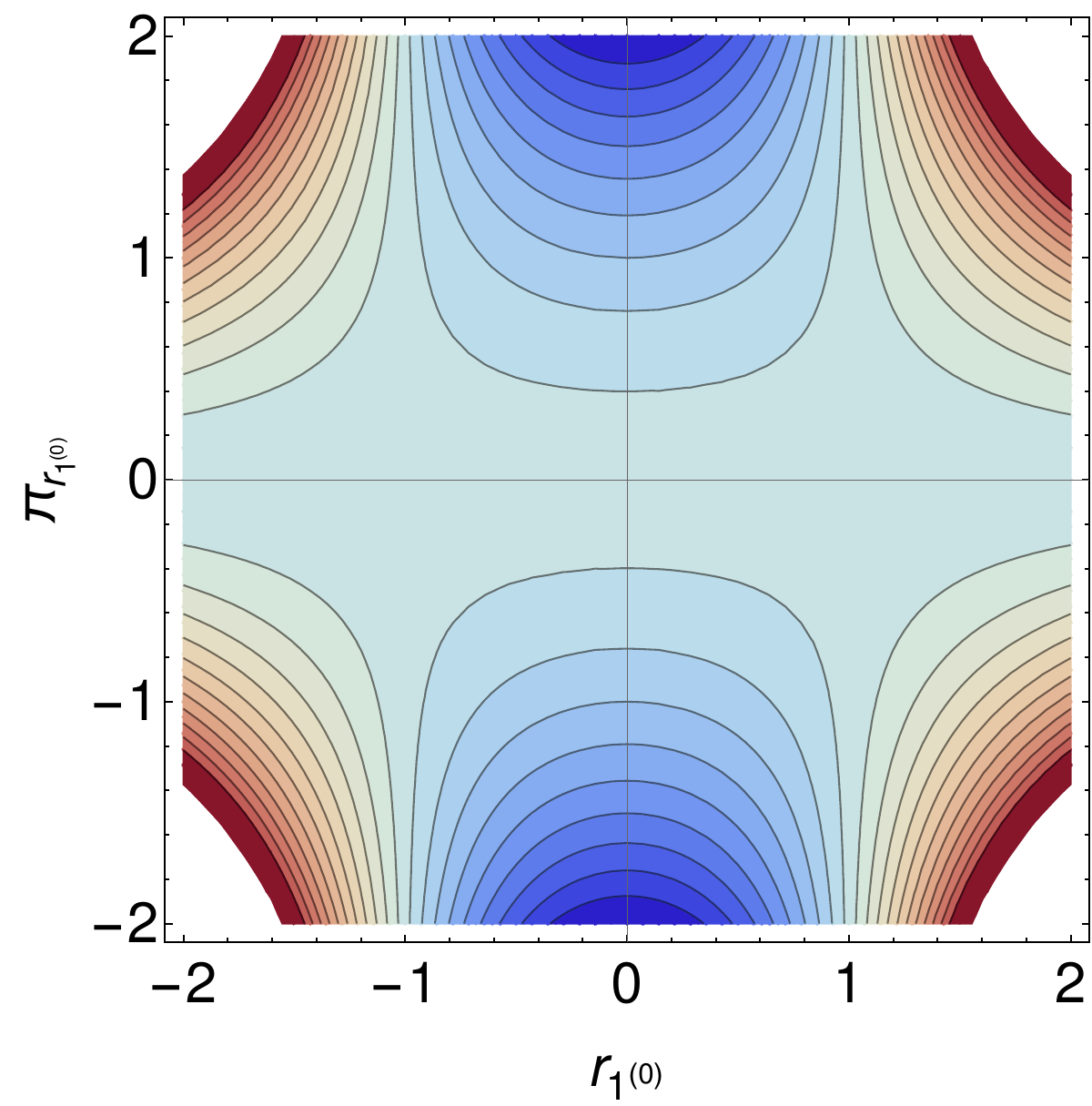} ~~\includegraphics[width=0.065\linewidth]{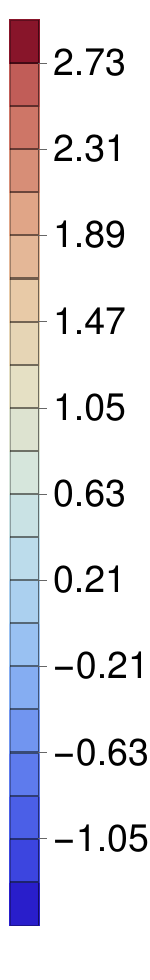}
     \caption{Phase space diagram for $\mathcal{H}_{LO}$ in the spinning configuration. The contour lines represent constant energy surfaces. The separatrix is given at ${r_1}^{(0)}=\pm 1$ separating the bounded (${r_1}^{(0)}< 1$) and unbounded (${r_1}^{(0)}>1$) region.}
     \label{Fig 1}
 \end{figure}

Quantization of the non-relativistic strings in curved target spaces are in general beyond the scope of this manuscript.
But in this section, to get an idea of how the energy spectra looks like, we will resort to semiclassics and discuss Bohr-Sommerfeld quantization for the models derived from LO/NLO terms of the worldsheet Lagrangian \cite{voros1976semi}. From the leading order Hamiltonian (\ref{LO Hamiltonian}), we have found periodic solutions for $r_1^{(0)}(\sigma)$ and $r_2^{(0)}(\sigma)$ already. Moreover, the canonical Hamiltonian does not have any explicit dependency on $\sigma$ and can serve as the conserved energy $E$ of the system, i.e., we can write
\begin{equation}
    E=\frac{1}{2}\left[e T_{eff}\kappa^2-\frac{1}{eT_{eff}}(\pi_{r_1^{(0)}})^2(1-(r_1^{(0)})^2)\right]
    \label{loe}
\end{equation}
From this equation, we may write
\begin{equation}
   (r_1^{(0)})^2\pi_{r_1^{(0)}}^2-\pi_{r_1^{(0)}}^2 =2EeT_{\text{eff}}\left(1-\frac{eT_{\text{eff}}\kappa^2}{2E}\right)
   \label{phase space equation}
\end{equation}The equation (\ref{phase space equation}) represents the geometry of the phase space of the corresponding solvable model in the $\pi_{r_1^{(0)}}$-$r_1^{(0)}$ plane. For finite positive effective tension $T_{\text{eff}}$, equation (\ref{phase space equation}) represents the phase space diagrams given in the Figure (\ref{Fig 1}). 
\medskip

Now let us elucidate a little on the semiclassical quantization for such systems.
Given a relativistic periodic system given by a coordinate $r_1(\sigma)$, the Bohr-Sommerfeld quantization condition can be considered as 
\begin{equation}
    \oint \pi_{r_1}dr_1=n\,,
    \label{relquan}
\end{equation}where, $n$ is the associated (effective) quantum number which labels different energy states. When we substitute the large $c$ expansion of the embedding $r_1$ and the corresponding canonical momentum $\pi_{r_1}$ for a non-relativistic framework, the condition (\ref{relquan}) gives 
\begin{equation}
      \oint \pi_{r_1^{(0)}}dr_1^{(0)}=n_{LO}\,,
      \label{LOosci}
\end{equation}at the leading order and 
\begin{equation}
      n_{NLO}=\oint \left[\pi_{r_1^{(0)}}dr_1^{(1)}+\pi_{r_1^{(1)}}dr_1^{(0)}\right]
      \label{NLOosci}
\end{equation}
at the NLO \footnote{This expansion can be visualized as $$n=\int \pi_{r_1} dr_1= \int (c^2\pi_{r_1^{(0)}}+ \pi_{r_1^{(1)}}+\mathcal{O}{(c^{-2})} )~d(r_1^{(0)}+1/c^2 ~r_1^{(1)}+\mathcal{O}{(c^{-4})}).$$ The arguments for this particular expansion series of longitudinal momenta can be found in \cite{Hartong:2021ekg}. Transverse momenta expansion is slightly different.}. Therefore, the Bohr-Sommerfeld quantization rule (\ref{LOosci}) for LO dynamics yields
\begin{equation}
    E=\frac{2(n_{LO})^2}{\pi^2eT_{eff}}+\frac{eT_{eff}\kappa^2}{2}
\end{equation}where, $n\in \mathbb{Z}_{+}$, and this is an exact relation. Evidently, the energies of different quantized states designated by $n_{LO}$ are discrete for the model obtained from LO dynamics. The threshold ground state energy is given by $E=\frac{eT_{\text{eff}}\kappa^2}{2}$ for $n_{LO}=0$ and then the other excited LO states have energies directly proportional to $n_{LO}^2$.
\\

Let us now focus on the case of the NLO dynamics and see if this can also be quantized in a similar way. Now, as the Hamiltonian (\ref{NRH}) does not explicitly depend on the $\sigma$, it can be considered as the conserved energy $E$ of the system.
Substituting for $\pi_{r_1^{(0)}}$ in terms of $E$ from the Hamiltonian (\ref{NRH}) and using $\pi_{r_1^{(1)}}=-eT_{eff}r_1'^{(0)}$ in the expression of the quantum number (\ref{NLOosci}), we get
\begin{equation}
\begin{split}
    n_{NLO}=E\int\frac{r_1'^{(1)}}{r_1'^{(0)}}d\sigma-&\frac{eT_{eff}}{2}\int \left(\frac{A^2}{(r_1^{(0)})^2}-\omega^2(r_1^{(0)})^2\right) \frac{r_1'^{(1)}}{r_1'^{(0)}}d\sigma\\&-eT_{eff}\int (r_1'^{(0)})^2d\sigma\,.
    \end{split}
\end{equation}One can then integrate the above expression over $0\leq\sigma\leq 2\pi$ by substituting exact solutions for the LO fields $r_1^{(0)}$ and $r_2^{(0)}$ and the LO onshell expression of $r_1^{(1)}$ given in (\ref{r11 eqn}). This yields
\begin{equation}
\begin{split}
E_{NLO}=\mathcal{A}~n_{NLO}+\mathcal{B}
    \end{split}
    \label{NLO E for spin}
\end{equation}
where, $\mathcal{A}$ and $\mathcal{B}$ are some nontrivial constants which depend on the values of $A,\omega,a$ and $T_{eff}$. As the explicit forms of these constants are quite intricate and also not enough illuminating in the context of our objective, we avoid writing those explicitly. However, the scaling (\ref{NLO E for spin}) is well-defined when $\cos \left(\frac{\pi  a}{2}\right)\geq 0\land \cos (\pi  a)\geq 0$ and $\sin \left(\frac{\pi  a}{2}\right)\geq 0\land \sin (\pi  a)\geq 0$. The take home message is, energies of the NLO quantized string energy levels scale linearly with $n_{NLO}$: 
\begin{equation}
    E_{NLO}\propto n_{NLO}
\end{equation} 
The threshold ground state energy is parameter dependent and designated by $\mathcal{B}(A,\omega,a, T_{eff})$. 
\section{Pulsating String in $\frac{1}{c^2}$-expanded sNC target space}
\label{section 5}
 In this section, we aim to extend our previous analysis to pulsating strings in the conformal gauge and derive the dispersion relation expressing the string energy in terms of the adiabatic oscillation number, considering both LO and NLO expansions as before. These solutions are very well known in the relativistic string literature, and are dual to particular operators in the usual gauge/gravity duality \cite{Minahan:2002rc}. They have radial temporal dynamics with the worldsheet embeddings oscillating in time.  To analyze the pulsating behaviour of the string, we start with an ansatz that mirrors the spinning one in \eqref{ansatzsp}:
\begin{equation}
    t^{(0)}(\tau)=\zeta \tau, ~ t^{(1)}(\tau,\sigma)=h(\sigma) \tau, ~ \theta^{(i)}(\tau,\sigma)=\theta^{(i)} (\tau),~\phi^{(i)}(\tau,\sigma)=m\sigma+f^{(i)}(\tau).
\end{equation}
Here $i=0,1$ corresponds to the LO and NLO coordinates, respectively. Further, $m$ is again an azimuthal winding number and $\zeta$ is a constant. Now let us introduce the embedding coordinates on the full (Lorentzian) sphere:
\begin{equation}
X_1+iX_2=\sin{\theta}e^{i\phi}=r_1(\tau)e^{i\Phi(\tau, \sigma)},~~ X_3=\cos{\theta}=r_2(\tau),~~t=t(\tau),~~\Phi(\tau,\sigma)=f(\tau)+m \sigma,
\end{equation}
Applying the $1/c^2$ expansion to the ansatz, 
the embedding coordinates can be expressed as before:
\begin{align}
    r_i(\tau)&=r_i^{(0)}(\tau)+c^{-2}r_i^{(1)}(\tau)+\mathcal{O}(c^{-4}),\\
   \Phi(\tau,\sigma)&=\Phi^{(0)}(\tau,\sigma)+c^{-2}\Phi^{(1)}(\tau,\sigma)+\mathcal{O}(c^{-4}),
\end{align}where, $r_1^{(0)}=\sin{\theta^{(0)}}$, $r_2^{(0)}=\cos{\theta^{(0)}}$, $r_1^{(1)}=\cos{\theta^{(0)}}\theta^{(1)}$ and $r_2^{(1)}=-\sin{\theta^{(0)}}\theta^{(1)}$
satisfy the constraint relation similar to \eqref{Constraints}, i.e, 
\begin{equation}
    (r_1^{(0)})^2+(r_2^{(0)})^2=1,~~r_1^{(0)}r_1^{(1)}+r_2^{(0)}r_2^{(1)}=0.
\end{equation}
Note that all coordinates are function of $\tau$ only. Further note here $m$ is the winding number, i.e. an integer, so we avoid performing a $1/c^2$ expansion of it. With the given background information, we now move on to analyzing the string dynamics at both LO and NLO.\\
\subsection{LO dynamics}
\label{subsection 5.1}
The Polyakov Lagrangian at leading order is given by \eqref{LOLag} which on substituting the LO pulsating string ansatz reduces to
\begin{align}
    \mathcal{L}_{LO}=-\frac{eT_{eff}}{2}\bigg(\zeta^2{-}\frac{(\dot{r}_1^{(0)})^2}{({r}_2^{(0)})^2}\bigg)=-\frac{eT_{eff}}{2}\bigg(\zeta^2{-}\frac{(\dot{r}_1^{(0)})^2}{1-({r}_1^{(0)})^2}\bigg),
    \end{align}
     The equation of motion for $r_1^{(0)}$ is given by
         \begin{equation}
        { \ddot{r}_1^{(0)}}+\frac{ r_1^{(0)}({\dot{r}_1^{(0)}})^2}{1-(r_1^{(0)})^2}=0.
         \end{equation}

     Given the initial condition $r_1^{(0)}(0)=0$, the solution to the above differential equation is given by $$r_1^{(0)}(\tau)=\sin({\alpha \tau}+\alpha_0)$$ where $\alpha$ is the integration constant.  This solution indicates that, at leading order,  the radial field exhibits the behavior of a simple harmonic oscillator. The energy and conjugate momenta corresponding to the leading order coordinates $t^{(0)}$ and $r_1^{(0)}$ respectively are given by,
     \begin{equation}
       E= -\frac{\partial \mathcal{L}_{LO}}{\partial \dot{t}^{(0)}}=eT_{eff}\zeta ,\quad \pi_{r_1^{(0)}}=\frac{\partial \mathcal{L}_{LO}}{\partial \dot{r}_1^{(0)}}=eT_{eff} \frac{\dot{r}_1^{(0)}}{1-(r_1^{(0)})^2}.
     \end{equation}
     Naturally, the canonical Hamiltonian of the system can be expressed as
     \begin{equation}
       \mathcal{H}_{LO}=\frac{1}{2}\left(eT_{eff} \zeta^2{+}\frac{(\pi_{r_1^{(0)}})^2(1-(r_1^{(0)})^2)}{(eT_{eff})}  \right)=E_{LO}.
       \label{Hamilton2}
     \end{equation}

\begin{figure}
          \centering
     \includegraphics[width=0.41\linewidth]{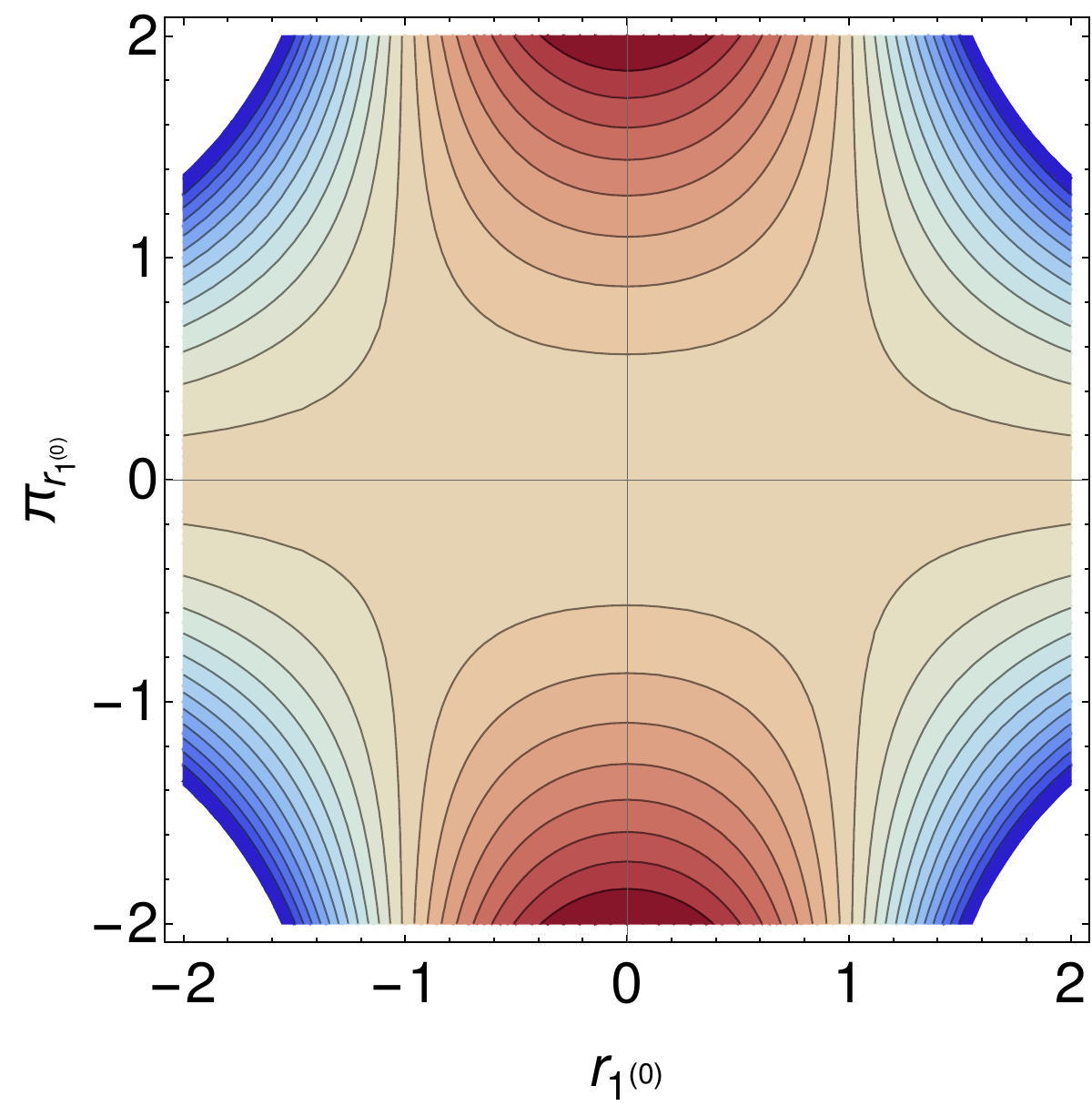} ~~\includegraphics[width=0.065\linewidth]{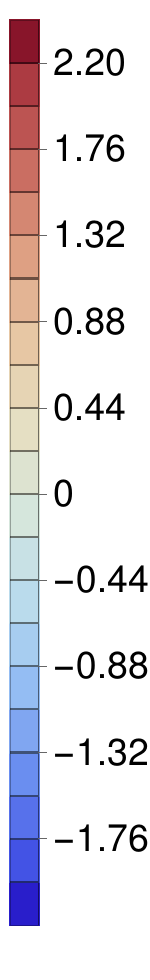}
         \caption{Phase space diagram for $\mathcal{H}_{LO}$ in the pulsating configuration. The contour lines represent constant energy surfaces. The separatrix is given at ${r_1}^{(0)}=\pm 1$ separating the bounded (${r_1}^{(0)}< 1$) and unbounded (${r_1}^{(0)}>1$) region.}
         \label{fig:4}
     \end{figure}

The phase space diagram of the model equivalent to LO pulsating string dynamics is similar to that achieved for LO spinning string dynamics as well, a plot of which can be found in Figure \eqref{fig:4}.
\medskip

     It is convenient to quantize the energies of a system of probe pulsating string by using the large adiabatic oscillation number. The oscillation number can then be  written by using the notion of the Bohr-Sommerfeld quantization principle as
         \begin{equation}
            \mathcal{N}_{LO}=T_{eff}\oint \pi_{r_1^{(0)}} dr_1^{(0)}.
         \end{equation}
       Substituting the momentum $\pi_{r_1^{(0)}}$ from \eqref{Hamilton2}, the oscillation number can be written as
   \begin{equation}
       \mathcal{N}_{LO} =T_{eff} \int_{0}^{1} \sqrt{\frac{2eT_{eff}E_{LO}-(eT_{eff}\zeta)^2}{1-(r_1^{(0)})^2}} dr_1^{(0)}
   \end{equation}
   After integration, we get the quantized energy levels of the LO system for pulsating string:
\begin{align}
 E_{LO}= \frac{2(\mathcal{N}_{LO})^2}{\pi^2eT_{eff}^3 }+\frac{eT_{eff}\zeta^2}{2}.
\end{align}

Therefore, it is evident here that, for a pulsating string, the energy of the leading order system increases with increasing oscillation number. The ground state threshold energy is $(E_{LO})_{\text{max}}=\frac{eT_{eff}\zeta^2}{2}$ and the energies of other quantized levels increase as $(\mathcal{N}_{LO})^2$. 

\subsection{NLO dynamics}
\label{subsection 5.2}
    At NLO, the Lagrangian density for a pulsating string, written with embedding radial variables, takes the form
    
    \begin{align}      
            \mathcal{L}_{NLO}={e T_{eff}}\Bigg[-\zeta h(\sigma)
    +\left(\frac{r_1^{(1)}\dot{r_1}^{(1)}+r_2^{(1)}\dot{r_2}^{(1)}}{\sqrt{(r_1^{(1)})^2+(r_2^{(1)})^2}}\right)\frac{{\dot{r}_1}^{(0)}}{\sqrt{1-(r_1^{(0)})^2}}+\frac{\left((\dot{f}^{(0)})^2-m^2\right)}{2}(r_1^{(0)})^2
     \Bigg].
     \label{Lpulsating}
         \end{align}
The equation of motion for $\dot{f}^{(0)}$ is given by 
  \begin{align}
        \dot{f}^{(0)}(\tau)=\frac{B}{(r_1^{(0)})^2},\quad
  \label{eomnlo}
\end{align}
where $B$ is a constant of integration. This reduces the Lagrangian (\ref{Lpulsating}) into
       \begin{equation}    
       \begin{split}
            \mathcal{L}_{NLO}={e T_{eff}}\left[- \zeta h(\sigma)+\left(\frac{r_1^{(1)}\dot{r_1}^{(1)}+r_2^{(1)}\dot{r_2}^{(1)}}{\sqrt{(r_1^{(1)})^2+(r_2^{(1)})^2}}\right)\frac{{\dot{r}_1}^{(0)}}{\sqrt{1-({r}_1^{(0)})^2}}\right. \\ \left.
 +\frac{1}{2}\left(\frac{B^2}{(r_1^{(0)})^2}-m^2(r_1^{(0)})^2\right)
     \right].
     \end{split}
         \end{equation}
 
\begin{center}
\begin{figure}[!]
\includegraphics[scale=0.38]{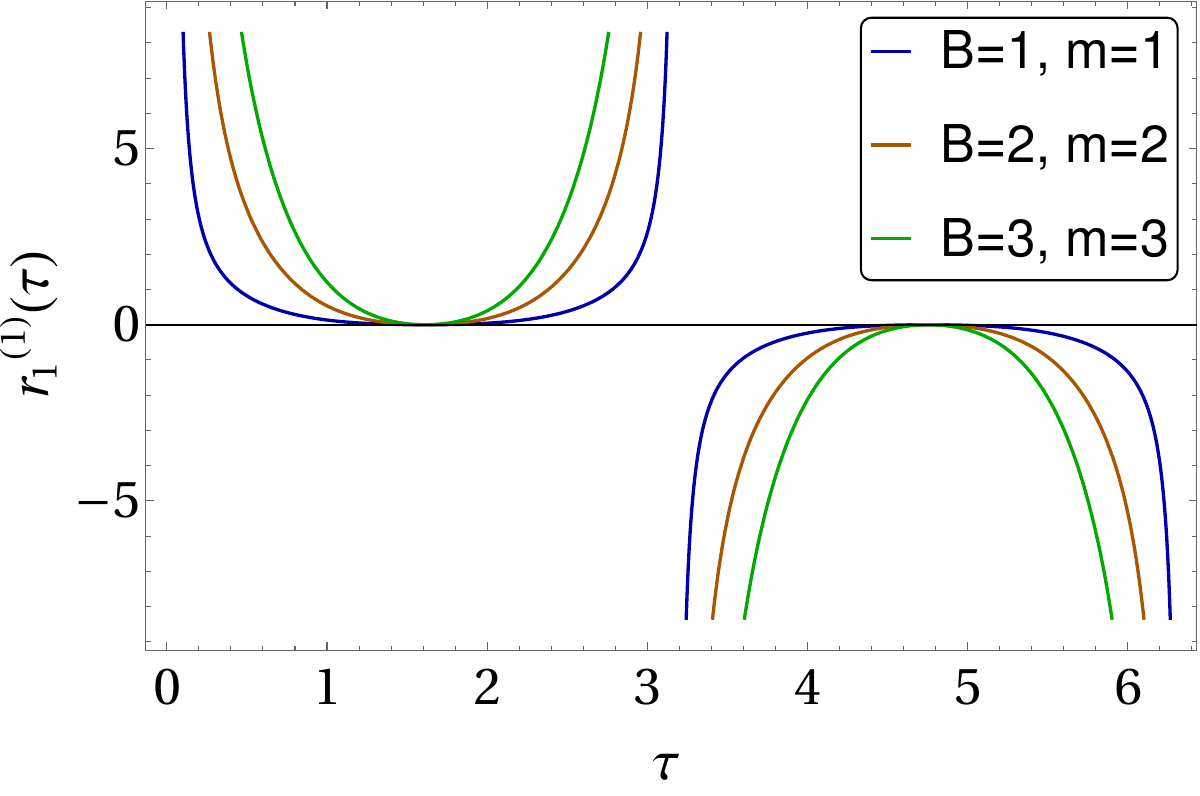}
\hfill
\includegraphics[scale=0.38]{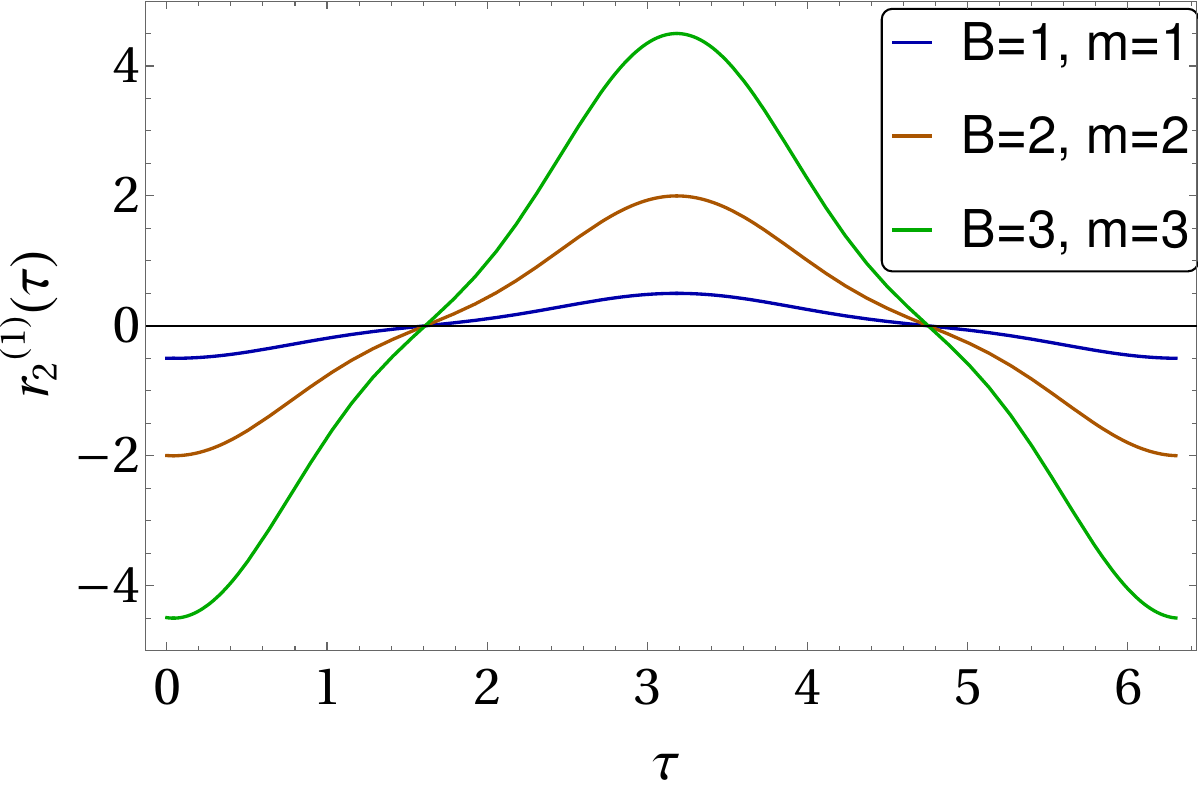}
\label{fig:2}
\caption{$r_1^{(1)}(\tau)$(left) and $r_2^{(1)}(\tau)$(right) for the NLO pulsating string solution. We take $\alpha=1$ and $\tau_0=3.1$, varying other parameters.}
\end{figure}
\end{center}

Similarly as done for the NLO dynamics of the spinning string, the equation of motion for $r_1^{(0)}$ gives 
\begin{equation}
    \theta^{(1)}(\tau)=\sqrt{(r_1^{(1)})^2+(r_2^{(1)})^2}= -\frac{B^2 \cot (\alpha  \tau +\tau_0)}{2 \alpha ^2}+\frac{m^2 \sin [2 (\alpha  \tau+\tau_0 )]}{8 \alpha ^2}
\end{equation}
Consequently we can reconstruct the solutions for NLO fields $r_1^{(1)}$ and $r_2^{(1)}$ as
\begin{equation}
    r_1^{1}(\tau)=\cos{(\alpha\tau+\tau_0)}\Big[-\frac{B^2 \cot (\alpha  \tau +\tau_0)}{2 \alpha ^2}+\frac{m^2 \sin [2 (\alpha  \tau+\tau_0 )]}{8 \alpha ^2}\Big]
\end{equation}
and
\begin{equation}
      r_2^{1}(\tau)=-\sin{(\alpha\tau+\tau_0)}\Big[-\frac{B^2 \cot (\alpha  \tau +\tau_0)}{2 \alpha ^2}+\frac{m^2 \sin [2 (\alpha  \tau+\tau_0 )]}{8 \alpha ^2}\Big]
\end{equation}
These solutions, obviously time periodic, are plotted in Figure \eqref{fig:2}.
Similar to the NLO analysis of the spinning string ansatz, we derive the canonical momenta for NLO pulsating string dynamics as
\begin{subequations}
\begin{align}
&\pi_{r_1^{(0)}}=eT_{eff}\left(\frac{r_1^{(1)}\dot{r}_1^{(1)}+r_2^{(1)}\dot{r}_2^{(1)}}{r_2^{(0)}\sqrt{(r_1^{(1)})^{2}+(r_2^{(1)})^{2}}}\right),\nonumber \\& 
   \pi_{r_1^{(1)}}=eT_{eff}\left(\frac{r_1^{(1)}\dot{r}_1^{(0)}}{r_2^{(0)}\sqrt{(r_1^{(1)})^{2}+(r_2^{(1)})^{2}}}\right)={eT_{eff}\dot{r}_1^{(0)}}\nonumber \\&
   \pi_{r_2^{(1)}}=eT_{eff}\left(\frac{r_2^{(1)}\dot{r}_1^{(0)}}{r_2^{(0)}\sqrt{(r_1^{(1)})^{2}+(r_2^{(1)})^{2}}}\right)={-eT_{eff}\frac{r_1^{(0)}}{r_2^{(0)}}\dot{r}_1^{(0)}},\nonumber \\&
   \pi_{t_1^{(1)}}=-eT_{eff}\dot{t}^{(0)},~~ \pi_{t_1^{(0)}}=-eT_{eff}\dot{t}^{(1)}.
\end{align}
\end{subequations}
Following exactly the rotating string scenario, the constraints between different orders of coordinates and momenta hold, the NLO dynamics of the pulsating string can also be reproduced from an equivalent Neumann-Rosochatius-like solvable model:
\begin{equation}
\begin{split}
    &H_{NR}=\frac{1}{e T_{eff}}\pi_{{r}_1^{(0)}}{\pi_{{r}_1^{(1)}}} -\frac{e T_{eff}}{2} \left[
     \frac{B^2}{(r_1^{(0)})^2}-m^2(r_1^{(0)})^2\right]\\&
    =\frac{1}{e T_{eff}}\left(\pi_{r_1^{(0)}}^{(1)}\pi_{{r}_1^{(1)}}-\frac{r_2^{(0)}}{r_1^{(0)}}\pi_{r_1^{(0)}}^{(2)}\pi_{{r}_2^{(1)}}\right) -\frac{e T_{eff}}{2} \left[
     \frac{B^2}{(r_1^{(0)})^2}-m^2(r_1^{(0)})^2\right] 
\label{HNLO}
\end{split}
\end{equation}where, $\pi_{r_1^{(0)}}=\pi_{r_1^{(0)}}^{(1)}+\pi_{r_1^{(0)}}^{(2)}$ and the condition $\pi_{r_1^{(1)}}r_2^{(1)}+\pi_{r_2^{(1)}}r_1^{(1)}=0$ still holds. 
The oscillation number at NLO can be expressed by
 \begin{equation}\mathcal{N}_{NLO}=T_{eff} \left[\oint \pi_{{r}_1^{(0)}} d{r}_1^{(1)} + \oint\pi_{{r}_1^{(1)}} d{r}_1^{(0)}\right]
 \label{oscillationnlo}
 \end{equation}As the Hamiltonian (\ref{HNLO}) does not explicitly depend on $\tau$, we can consider it as the total energy $E_{NLO}$ of the equivalent solvable model for the NLO pulsating string dynamics. Using the expression \eqref{HNLO} and working in the same vein as in the spinning string case, i.e.
substituting all the coordinates along with their derivatives in terms of the chosen ansatz, the expression in (\ref{oscillationnlo}) reduces to
\begin{equation}
\begin{split}\mathcal{N}_{NLO}=\mathcal{P}+E_{NLO}\mathcal{Q}
    \end{split}
    \label{NLO N for puls}
\end{equation}where, $\mathcal{P}$ and $\mathcal{Q}$ are again some nontrivial constants which depend on the values of $\alpha, \tau_0, B, m$ and $T_{eff}$. Similarly as in the spinning string case, here also the expression (\ref{NLO N for puls}) is well-defined with these constants only for $\cos (\tau_0)\geq 0\land \cos \left(\tau_0+\frac{\pi  \alpha }{2}\right)\geq 0$ and $\sin (\tau_0)\geq 0\land \sin \left(\tau_0+\frac{\pi  \alpha }{2}\right)\geq 0$. Thus the energies of the NLO pulsating string states are found to be directly proportional to $\mathcal{N}_{NLO}$ as
\begin{equation}
    E_{NLO}=\tilde{\mathcal{Q}}\mathcal{N}_{NLO}+\tilde{\mathcal{P}},~~\tilde{\mathcal{Q}}=\frac{1}{\mathcal{Q}},~~\tilde{\mathcal{P}}=-\frac{\mathcal{P}}{\mathcal{Q}}\end{equation}The threshold groundstate energy for $\mathcal{N}_{NLO}$ is specified by $E_0=\tilde{\mathcal{P}}(\alpha, \tau_0, B, m, T_{eff})$. One must compare this semiclassical LO/NLO energy with the relativistic case of a ``short'' pulsating string ($\mathcal{N}\to 0$) on two sphere, where the energy scales like $\mathcal{N}^{1/2}$ in the leading order \cite{Beccaria:2010gu}. 
 \section{Summary and Conclusion}
 \label{section 6}
 \subsection*{This work}
In this paper, we embarked upon the quest to understand classical string solutions and solvable structures thereof, when the target space is a non-relativistic version of $\mathbb{R}\times S^2$. Exact string solutions for non-relativistic string sigma model in curved spaces are scarce in the literature, and we discussed various classes of simple embeddings in our case. We used a two-pronged approach, studying the problem both from the intrinsic sNC model, and then reinforcing it from a $1/c^2$ expansion of the relativistic string sigma model.
\medskip

We discussed GKP and spinning strings in the intrinsic model, finding exact solutions despite the non-trivial constraint structure. We also found the energy-spin dispersion relations for the same, and tried to interpret those from our intuition of relativistic holography. From the same intuition in the semiclassical limit, i.e. in regimes where charges scale with large tension, these dispersions should give one information on the anomalous dimensions of putative dual operators. Remarkably, we found in the sNC intrinsic sigma model, the GKP-like solution still holds a similar dispersion relation as its relativistic parent. On the other hand, the spinning string seemingly gives rise to a small momentum limit of the Giant Magnon relation. 
\medskip

We further went ahead and discussed consistent $1/c^2$ expansions of relativistic string sigma model on the two sphere. From resulting LO and NLO actions, we found solutions for both the spinning and pulsating string embeddings. We subsequently reduced these actions to those akin to a Neumann-Rosochatius dynamical system. We ended our discussion with an attempt at semiclassical quantization for the energy of spinning/pulsating string, as function of certain oscillation numbers. The corresponding dependence of $E\sim f(\mathcal{N})$ turns out to be significantly different from relativistic cousins, further igniting our intrigue. However, without a clear idea of dual operator spectrum, it is hard to speculate on the nature of such solutions. 

 \subsection*{The road ahead}

 Going forward, there are various avenues one can explore in the regime of non-relativistic string solutions and related exploration of integrable structures. As we mentioned previously, the community is only gearing up with the study of non-relativistic strings, and some immediate but solid goals can easily be set.
 \medskip

To start with, the string solutions we considered in this work, can give us some more hints of the integrable structures. For example, for the spinning string solutions we considered in \eqref{section 2.3}, since it is a soliton solution, one can write down an associated sine-Gordon model \cite{Chen:2006gea}. These solutions can be related via transforming the action using what is known as \textit{Pohlmeyer reduction} \cite{Pohlmeyer:1975nb}. The complex sine-Gordon solitons with two charges can be mapped to \textit{Dyonic Giant Magnons}, which can be found by considering similar solutions in $\mathbb{R}\times S^3$. An intrinsic sNC version of this computation should be imminent, however subtle. If this mapping in this case takes us to a known hierarchy of integrable theories, one can exploit the relevant structure wisely, even maybe for constructing multi-soliton solutions. 
\medskip

A similar exercise should be doable in sNC cousin of $AdS$  backgrounds as well. String solitons there, like the large spin GKP solution (or \textit{Spiky strings} \cite{Kruczenski:2004wg}), can be mapped to the sinh-Gordon integrable field theories. Related field theory reduction and inverse scattering techniques for the relativistic case have been described in \cite{Jevicki:2007aa} and numerous follow-ups. If one can map classical solitons from sNC sigma model to an integrable field theory setting, the bigger picture should be reasonably clear \footnote{In the usual relativistic string theory, this mapping was done by the 90's as Larsen, de Vega and Sanchez with collaborators were able to show string theory on $AdS_2$, $AdS_3$ and $AdS_4$ is equivalent to
Liouville theory, sinh-Gordon theory and $B_2$ Toda theories respectively \cite{DeVega:1992xc, Combes:1993rw, Larsen:1996gn}. A classification like this may be tempting but difficult in the sNC case.}. This is something we will be interested in very near future.
\medskip

 There are of course other related progress. It is known that certain non-Lorentzian strings with a Galilean worldsheet (like worldsheet non-relativistic limit of the Gomis–Ooguri string) are conjectured to be dual to \textit{Spin Matrix Theory} (SMT)\cite{Harmark:2017rpg, Harmark:2018cdl}, which arises as a near-BPS limit of $\mathcal{N}=4$ super Yang–Mills theory. This is actually a very powerful tool that provides an exact map between non-relativistic sectors of relativistic string theory and the duals in holography \cite{Harmark:2014mpa}. Some classical strings have also been studied in this setup \cite{Roychowdhury:2020yun}, and more detailed investigations seem like a real possibility.
 \medskip
 
Also, much of the focus on various non-Lorentzian stringy constructions has been dominated by closed strings. Although recently \cite{Hartong:2024ydv} the $1/c^2$ expansion bosonic open relativistic sigma model has been performed. Different boundary conditions and D-brane-like states have also been discussed. This opens a whole new arena to study classical solutions and related integrable structures in this regime.
\medskip

The field of non-Lorentzian classical strings and algebraic structures associated to them is mostly at its infancy, and there are plenty of adventures to be had in these realms. We hope to get back to some of these issues in future correspondences. 

\acknowledgments
ArB is supported in part by an OPERA grant and a seed grant NFSG/PIL/2023/P3816 from BITS-Pilani and an early career research grant ANRF/ECRG/2024/002604/PMS from ANRF. He also acknowledges financial support from the Asia Pacific Center for Theoretical Physics (APCTP) via an Associate Fellowship.
AC would like to thank the  Research project supported by program "Excellence initiative – research university" for the AGH University under the project IDUB 501.696.7996 for providing funds to carry out the above work. 
\appendix
\section{Relativistic Neumann-Rosochatius system}
\label{appendix A}
The classical Neumann-Rosochatius model is an integrable model that depicts the motion of a one-dimensional harmonic oscillator constrained on a $N$-dimensional unit sphere under the influence of an extra inverse-square centrifugal potential. Such integrable model has been established as an equivalent classical framework for studying a large class of generic spinning and pulsating string solutions as the worldsheet Lagragians of such strings while probing any chosen 10D AdS/S compact target space reduce into that of Neumann-Rosochatius model. 
\subsection*{The dynamical system}
Let us consider $\mathbb{R}\times S^2$ as the simplest subspace of any such target space. When we probe a usual  fundamental string in the relativistic $\mathbb{R}\times S^2$ background, the ansatz for rigid closed spinning and pulsating string result in the worldsheet Lagrangian given as 
\begin{equation}
    L=\frac{T}{2}\left[r_1'^2+r_2'^2+\frac{A^2}{r_1^2}-\omega^2r_1^2+\frac{\Lambda}{2}\left(r_1^2+r_2^2-1\right)\right]
    \label{relLforspin}
\end{equation}and
\begin{equation}
    L=\frac{T}{2}\left[\dot{r}_1^2+\dot{r}_2^2+\frac{B^2}{r_1^2}-m^2r_1^2+\frac{\Lambda}{2}\left(r_1^2+r_2^2-1\right)\right] 
    \label{relLforpuls}
\end{equation}respectively. It is evident that, for both of these generic string ansatz, the Lagrangian for the relativistic string dynamics contain both Harmonic-oscillator-type and inverse-square-type potentials and follows the constraint of the geometry of a unit 2D sphere, constraint being suitably incorporated by using Lagrange multiplier $\Lambda$. Hence, such systems of relativistic probe spinning and pulsating strings in $\mathbb{R}\times S^2$ can be equivalently described by the construction of the classical Neumann-Rosochatius model. For spinning string, the equation of motion for the embedding $r_1(\sigma)$ gives
\begin{equation}
    r_1^{''}+\frac{A^2}{r_1^3}+\omega^2 r_1=0
\end{equation}with $A$ being a constant of integration and $\omega$ is the constant frequency. This generates the solution as
\begin{equation}
    r_1(\sigma)=\text{sn}(\eta \sigma|k)
    \label{relsolspin}
\end{equation}where, $\eta$ and $k$ are some constants involving $A$ and $\omega$. Similarly, for a pulsating string, the equation of motion for $r_1(\tau)$ can be derived from the Lagrangian (\ref{relLforpuls}) as
\begin{equation}
      \ddot{r}_1+\frac{B^2}{r_1^3}+m^2 r_1=0
\end{equation}This also yields
\begin{equation}
     r_1(\tau)=\text{sn}(\tilde{\eta} \tau|\tilde{k})
    \label{relsolpul}
\end{equation}Here, the constants $\tilde{\eta}$ and $\tilde{k}$ involve $B$ and $m$. The above models are both integrable as they consist of the standard form of the Uhlenbeck integrals of motion of a classical Neumann-Rosochatius model, and those integrals of motion are found to be in involution. The solutions of such an integrable framework generate a BMN-like dispersion relation $E\propto J$ between the conserved energy and angular momentum of spinning string states for large $J$ limit. For the pulsating string case, the solutions of integrable NR framework yield the energies as $E\propto\sqrt{\mathcal{N}}$, where $\mathcal{N}$ denotes the large adiabatic oscillation number. 

\subsection*{Integrals of motion}

Our approach to constructing the integrals of motion is to first determine the relativistic integrals of motion in $\mathbb{R} \times S^2$ and then perform a $1/c^2$ expansion of these integrals. Let's begin with a relativistic spinning string moving in $\mathbb{R} \times S^2$. The corresponding Lagrangian is given as 
\begin{equation}
    \mathcal{L}=-\frac{T}{2}\left[\dot{t}^2+\theta'^2+\sin^2{\theta}(\phi'^2-\dot{\phi}^2)  \right].
\end{equation}
Expressing this in terms of the embedding coordinates introduced in equation(\ref{embed2}),  the system can be expressed as 
\begin{equation}
    \mathcal{L}=-\frac{T}{2}\left[\dot{t}^2+\frac{r_1'^2}{1-r_1^2}-\omega^2r_1^2+\frac{A_1^2}{r_1^2}\right],
\end{equation}
 The equation of motion for $r_1$ becomes
\begin{equation}
\frac{r_1''}{1-r_1^2} +\frac{r_1 {r_1'}^2}{\left(1-r_1 ^2\right)^2}+\frac{A^2}{r_1 ^3}+\omega^2r_1=0.
\end{equation}
which can also be followed from a Neumann-Rosochatius-like system given by
\begin{equation}
   {L}=\frac{1}{2}\left[\frac{r_1'^2}{(1-r_1^2)}-\omega^2r_1^2+\frac{A_1^2}{r_1^2}\right].
\end{equation}
The Hamiltonian of the system is given by
\begin{equation}
     {H}=\frac{1}{2}\left[\pi_{r_1}^2 (1-r_1^2)+\omega ^2r_1^2 -\frac{A_1^2}{r_1^2}\right],
\end{equation}
where $\pi_{r_1}=\frac{{r_1}'}{{1-r_1}^2}$.
The Uhlenbeck integral of motion for such a system can be expressed by \cite{Arutyunov:2003uj}
\begin{equation}
   I=r_1^2+\frac{1}{\omega^2}\left[\frac{{r_1'}^2}{{1-r_1}^2}+\frac{A_1^2r_2^2}{r_1^2}\right].
\end{equation}
which satisfies the Poisson commutation relation with the Hamiltonian of the system, i.e., $\{I,{H} \}=0$ with the constraint $r_1^2+r_2^2=1$.

\section{Comments on non-relativistic integrals of motion}\label{appendix C}
The equivalent solvable models obtained at the LO and NLO dynamics of the large $c$-expansion acquire the Neumann-Rosochatius forms of Lagrangian and Hamiltonian. At this stage, it is straightforward to look into the probable Liouville integrability present in these dynamics. In general, 
the integrability of the Neumann-Rosochatius system follows from the existence of a set
of integrals of motion in involution, the so called Uhlenbeck invariants \cite{10.1007/BFb0069763}. In what follows we attempt a systematic anatomization of the integrals of motion as well as their Poisson bracket involution to study the Liouville integrability of both of the LO and NLO dynamics. 

\subsubsection*{LO Integrals of Motion:}
The leading order NR-like Hamiltonian of the system can be extracted from the spinning string LO Hamiltonian \eqref{LO Hamiltonian}, and is given by
\begin{eqnarray}
\begin{split}
   H_{LO}=\pi_{r^{(0)}_1}^2(1-{r^{(0)}_1}^2).
     \end{split}
 \end{eqnarray}
 Now the conformal constraint gives us $ \mathcal{H}_{LO}=\kappa^2$.
Implementing this, we obtain
  \begin{equation}
    {r'^{(0)}_1}^2=\kappa^2  ({r^{(0)}_2}^2).
  \end{equation}
 The integral of motion is given by
 \begin{equation}
    I= ({r^{(0)}_2})^2.
 \end{equation}The system is Liouville integrable as the integral of motion satisfies $\Big\{I,\mathcal{H}_{LO}\Big\}=0$. Thus, the exactly solvable model found at the leading order dynamics of the large $c$-expansion is Liouville integrable.
\subsubsection*{NLO Integrals of Motion:}
The effective Lagrangians at NLO for both rotating and pulsating string cases appear in the forms similar to that of a Neumann-Rosochatius type exactly solvable model. 
Nevertheless, due to the large $c$ expansion, the kinetic term in the Lagrangian undergoes a deformation which in turn affects the integrals of motion. To account for this deformation, we extend the standard Uhlenbeck integrals of motion for the Neumann-Rosochatius model with an additional function $f(r_1^{(1)},\pi_{r_1}^{(1)})$.
\begin{equation}
    I=(r_1^{(0)})^2+\frac{1}{\omega^2}\left[ (r_1^{(0)}\pi_{r_1^{(1)}}-r_1^{(1)}\pi_{r_1^{(0)}})^2 +\frac{B^2}{(r_1^{(0)})^2}(r_1^{(1)})^2\right]+f(r_1^{(1)},\pi_{r_1^{(1)}}).
    \label{NLOIOm1}
\end{equation}
To consider the above expression as the integral of motion of the equivalent NR-like model at NLO, we must need the condition
\begin{equation}
    I'=0.
    \label{C.10}
\end{equation} After substituting the solutions for the $r$-coordinates of both the LO and NLO dynamics in (\ref{NLOIOm1}) and using the condition (\ref{C.10}), we get with LO on shell ,
\begin{equation}
    \begin{split}
    f(\sigma)=-\frac{1}{64 a^3 \omega ^2}\left[-8 e T_{eff} \left(4 a^4+A^2 \omega ^2\right) \sin (2 (a\sigma+b))\right. \\ \left.+16 a \cos (2(a\sigma+b)) \left(B^2 \omega_0^2 \left(\omega_0^2-4 A^2\right)-2 a^2 \omega ^2\right)\right. \\ \left.+128 a A^4 B^2 \csc ^4(a\sigma+b)+8 A^2 \cot (a\sigma+b) \left(e T_{eff} \omega ^2+\omega_0^2\right)\right. \\ \left.+128 a A^2 B^2 \left(\omega_0^2-2 A^2\right) \csc ^2(a\sigma+b)+4 a B^2 \omega_0^2 \left(16 A^2-5 \omega_0^2\right)\right. \\ \left.+16 A^2 \cot (a\sigma+b) \csc ^2(a\sigma+b)+4 a B^2 \omega_0^4 \cos (4 (a\sigma+b))\right. \\ \left.+\omega_0^2 \omega ^2 \sin (4(a\sigma+b))\right]
    \end{split}
\end{equation}Note that, we derive the deformed part as a function of $\sigma$ as we replaced all the position and momenta coordinates with their respective LO on shell expressions. Substituting this deformation in (\ref{NLOIOm1}), we get the deformed integral of motion as 
\begin{equation}
\begin{split}
    I=\frac{1}{16 a^2 \omega ^2}\Big[\left(8 a^2-1\right) B^2 \left(4 \omega_0^2 \left(\omega_0^2-4 A^2\right) \cos (2 (a\sigma+b))\right. \\ +32 A^2 \csc ^2(a\sigma+b) \left(A^2 \csc ^2(a\sigma+b)-2 A^2+\omega_0^2\right) \\ +\omega_0^4 \cos (4(a\sigma+b))+8 a^2 \omega ^2 \\ +B^2 \left(256 a^2 A^4-16 \left(24 a^2+1\right) A^2 \omega_0^2+\left(24 a^2+5\right) \omega_0^4\right)\Big]
\end{split}
    \label{deformed IOM}
\end{equation}
Again, for an integrable physical system we must have $\Big\{I,H_{NR}\Big\}=0$. If one writes all the trigonometric functions in (\ref{deformed IOM}) in terms of the leading order solution, then Poisson commutation occurs only with the condition 
\begin{equation}
    -\frac{1}{2}\omega_0^2(\omega_0^2-4A^2)(r_1^{(0)})^2+\frac{2A^4}{(r_1^{(0)})^4}+\frac{2A^2}{(r_1^{(0)})^2}(\omega_0^2-2A^2)+\frac{\omega_0^2}{8}(r_1^{(0)})^2\left(4(r_1^{(0)})^2-3\right)=\text{Const.}
\end{equation}

\bibliographystyle{JHEP}
\bibliography{NRstring}
\end{document}